\documentclass[12pt]{article}
\usepackage{amsfonts}
\usepackage{amssymb}
\usepackage{epsfig}
\usepackage{graphicx}
\usepackage{dcolumn}
\usepackage{bm}
\usepackage{lscape}
\def\Vec#1{\mbox{\boldmath $#1$}}
\def\eqne{\end{equation}}
 
\def\eqnb{\begin{equation}}
\def\PTP{Prog. Theor. Phys.(Kyoto)}

\def\NPB{{Nucl. Phys.} {\bf B}}
\def\PLB{{Phys. Lett.} B}
\def\PRL{Phys. Rev. Lett.}
\def\PRD{{Phys. Rev.} D}

\def\FBS{Few-Body Systems}

\title{Fermion flavors in quaternion basis and infrared QCD}
\author{Sadataka Furui \\
Faculty of Science and Engineering, Teikyo University.\\
1-1 Toyosatodai, Utsunomiya, 320-8551 Japan\thanks{\textit{E-mail address:} furui@umb.teikyo-u.ac.jp }}

\begin{document}

\maketitle

\begin{abstract}
I analyze the lattice simulation data of the Domain Wall Fermion in quaternion basis. 
As pointed out by Atiyah and Ward, the minimum action solution for SU(2) Yang-Mills fields in Euclidean 4-space correspond, via Penrose twistor transform, to algebraic bundles on the complex projective 3-space. Assuming dominance of correlation between the fermions on the domain walls via exchange of instantons, I extract parameters necessary for defining gauge fields of Atiyah-Ward ansatz.  
The QCD effective coupling in the infrared and the relation between the number of flavors and the infrared fixed point is investigated.
  
Consequence of this lepton flavor assignment to phenomenology of baryons is also discussed.
\end{abstract}

\section{Introduction}
The properties of infrared QCD depend on the flavor number of fermions.
In 1982, Banks and Zaks\cite{BZ82} showed that the 3-loop $\beta$ function of an SU(3) gauge theory 
with $N_F$ Dirac fermions in the representation R is
\begin{eqnarray}
\beta(g)&=&-\left(\beta_0\frac{g^3}{16\pi^2}+\beta_1\frac{g^5}{(16\pi^2)^2}+\beta_2\frac{g^7}{(16\pi^2)^3}\right)\nonumber\\
\beta_0&=&11-\frac{4}{3}T(R)N_F\nonumber\\
\beta_1&=&102-(20+4C_2(R))T(R)N_F\nonumber\\
\beta_2&=&(\frac{285}{2}-\frac{5033}{18}N_F+\frac{325}{54}N_F^2)
\end{eqnarray}
From the ratio of $\beta_0$ and $\beta_1$ which are invariant under the renormalization scheme,
they showed that there appears an infrared fixed point when $N_F=\frac{11N_c}{2}=16.5$

Using the staggered fermion and the Schr\"odinger functional scheme, Appelquist et al\cite{AFN08,App09} claimed that if $N_f$ is in the range of $12\leq N_f\leq 16$, the infrared behaviors are governed by the IR fixed point.
In the $N_f=4$ system, the step scalng function shows size dependence whe the coupling constant $u$ is large\cite{App09, RS10}. Since staggered fermion has the 4 times taste degeneracy, the interpretaion of the effective $N_f$
is not simple. Recently, Fodor et al. \cite{FHKNS11} showed that the dependence of chiral condensates and the
pion decay amplitude $F_\pi$ do not show the expected dependence on the QCD anomalous dimension $\gamma$ 
and cast doubt on the conformal window hypothesis. Staggered fermion do not have Dirac $\gamma$ matrices and
the algebraic structure could be diffrent from that of Wilson fermion.

Dietrich and Sannino\cite{DS07} studied critical flavor number for a presence of Banks-Zaks type fixed point  in two-loop $\beta$ function for a generic non-Abelian gauge theory with fermionic matter in a given representation of SU(N).  Non-perturbative corrections to these perturbative theory is not clear. 

Grunberg\cite{Gr94,GG01,Gr01}, assumed that there is two-phase structure in QCD as the number of flavors is changed.
For $N_f^*< N_f<N_f^0=16.5$, the conformal field amplitude $D(Q^2)$ is calcuated from perturbation theory  $D_{PT}(Q^2)$ and in the region $N_f<N_f^*$ the non-perturbative correction $D_{np}(Q^2)$ is to be added. It is also
assumed that there is non-trivial negative UV fixed point $\alpha_{UV}$. 
The $\beta$ function which allows non-trivial  $\alpha_{UV}$ and $\alpha_{IR}$ is expressed as
\begin{equation}
\tilde\beta(\alpha)=-\beta_0\alpha^2+\tilde\beta_1\alpha^3-\cdots,
\end{equation} 
where $\tilde\beta_1=\rho\beta_0$. The parameter $\rho$ is adjusted such that $\tilde\beta(\alpha)=0$
for $\beta_0=0$ i.e. $N_f=16.5$ and for each $N_f$
\begin{equation}
 \tilde \beta(\alpha)=-\beta_0\alpha^2+\beta_1\alpha^3
\end{equation}coincides with the true $\beta(\alpha)$.  He obtained $\alpha_{IR,UV}$ as
\begin{equation}
\tilde \alpha_{IR,UV}=\pm\tilde\epsilon(1+\frac{1}{2}\frac{\beta_{30}}{\beta_{20}}\tilde\epsilon+\cdots)
\end{equation}
where $\tilde\epsilon=(-\frac{\beta_0}{\beta_{20}})^{1/2}$ and $\beta_{i0}$ are the $\beta_i$ at $N_f=16.5$.
In this modified Banks-Zaks fixed point theory, the conformal window is predicted in $4\leq N_f\leq 6$              

The resummed running coupling $R_0(Q^2)$ is defined as
\begin{equation}
R_0(Q^2)=\int_0^\infty a(k^2)\phi(k^2/Q^2)\frac{dk^2}{k^2}
\end{equation}
where $a(k^2)=\frac{\alpha_s(k^2)}{\pi}$ and $\phi(k^2/Q^2)$ is observable dependent Feynman integrand.

Singulaity  of the Borel transform of $R_0$ defined as 
\begin{equation}
R_0(Q^2)=\int_0^\infty B(z)e^{-z/a(Q^2)}dz
\end{equation}
on the real axis in the $z$ complex plane are called renormalons. One assumes
\begin{equation}
B(z)=\frac{1}{[1-(z/z_p)]^{1+\delta}}
\end{equation}
where $\delta=\frac{p\beta_1}{\beta_0^2}$ and $z_p=p/\beta_0$ is the renormalon location. Banks-Zaks expansion  with infrared fixed points has factorial divergence.

Although it is not a direct evidence, infrared freezing of the QCD effective coupling is an indication
 of proximity of the system to the conformal window.
In MOM scheme, the quark-gluon coupling $\alpha_s(q)$ in the Coulomb gauge of the 2+1 flavor domain wall fermion (DWF)\cite{DWF07}, and the ghost-gluon coupling $\alpha_s(q)$\cite{FN07b} are calculated.  In Fig.1, the results are compared with the experimental extraction of $\alpha_{g1}(q)$ derived from the polarized electron proton scattering by the JLab group\cite{DBCK06}.

\begin{figure}
\begin{minipage}[b]{0.47\linewidth}
\begin{center}
\includegraphics[width=6cm,angle=0,clip]{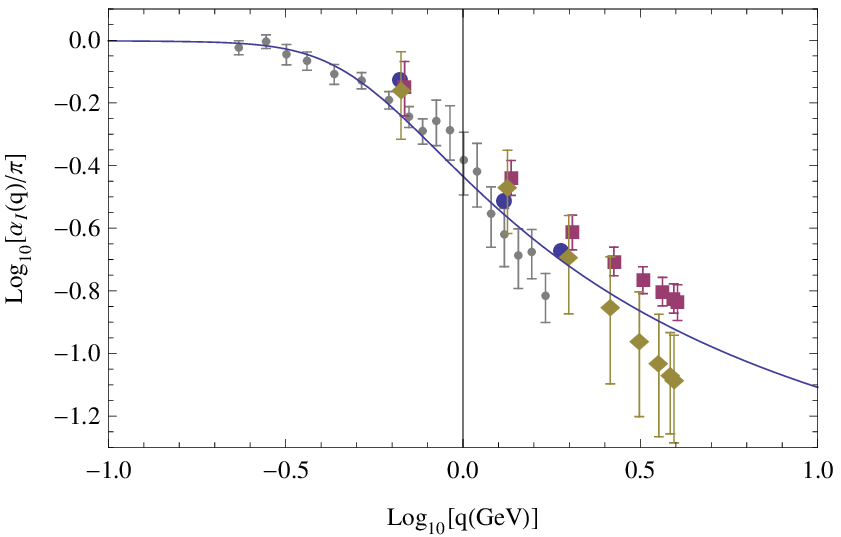}
\caption{The effective coupling of QCD. Jlab experiment and lattice simulation.
Small disks from $\log_{10}q$ from 0.5 to 0.3 and the solid curve are the $\alpha_{g1}(q)$ of JLab. 
Diamonds and squares are $\alpha_s(q)$ of ghost-gluon coupling of quark mass $m_u=0.02/a$ and $m_u=0.01/a$, respectively\cite{FN07b}.
Three large blue disks are $\alpha_s(q)$ of the quark-gluon coupling. }
\label{alp_jlab}
\end{center}
\end{minipage}
\hfill
\begin{minipage}[b]{0.47\linewidth}
\begin{center}
\includegraphics[width=6cm,angle=0,clip]{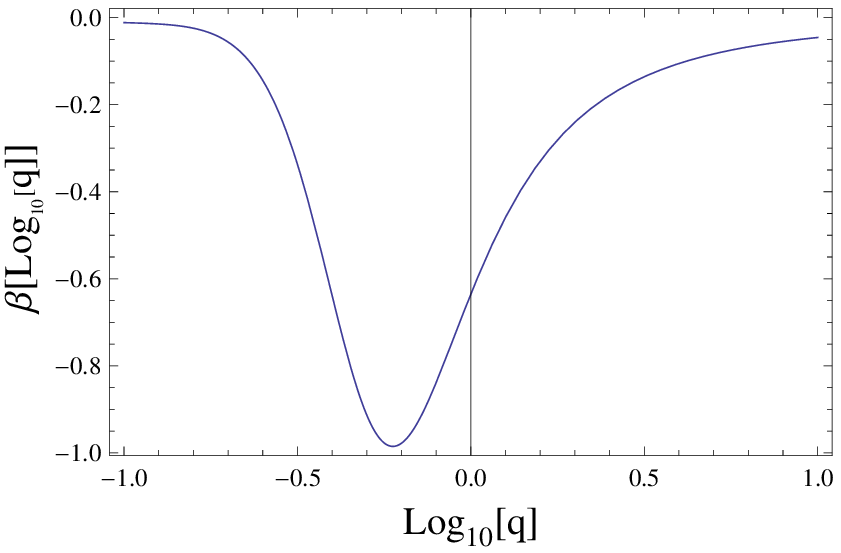}
\caption{The beta function obtained from the fitted function in the Figure 1.}  
\label{beta_jlab}
\end{center}
\end{minipage}
\end{figure}

The fitted line of the $\alpha_{g1}(q)$ is derived from \cite{DBCK06} and the corresponding beta function is calculated numerically using Mathematica\cite{wolfram}, which is shown in Fig.\ref{beta_jlab}. In \cite{BTD10}, $\alpha_s^{AdS}(q)$ of $AdS_5\times S^5$ \cite{TB05} theory at high energy and the phenomenological $\alpha_{g1}(q)$ at low energy are combined. In the infrared, the correction from $\alpha_s^{AdS}$ is small.
The coupling constant in different scheme for calculating a physical quantities are related by commensurate scale relation\cite{BLM83}. It is remarkable that $\alpha_s^{AdS}(q)$ is close to $\alpha_{g1}(q)$.

The beta function does not have zero, probably due to the isometry of AdS space mapping, $x_\mu\to \lambda x_\mu$ and the holographic coordinate space mapping $z\to \lambda z$. 

The freezing of the running coupling  is observed also in  $\alpha_\tau(s)$ in the infrared\cite{BMMR03}. In this work,  essentially only the experimental data of $\tau$ decay was used.
 Beneke and Braun\cite{BB94} pointed out crucial roles played by renormalons in the Brodsky-Repage-Mackenzie(BLM) prescription of scale fixing.  In BLM the scale changing to $Q^*$ by incorporation of fermion loops has the same effect as resummation of diverging radiative corrections.
The infrared freezing of the running coupling observed by  Brodsky et al could be understood in the same framework as Grunberg's effective charge theory\cite{BGGR01} . 

  Brodsky and Shrock\cite{BS08} discuss low energy QCD phenomenology and  point out that instanton effect may be suppressed by a presence of confinement induced gluon momentum $k_{min}$

 Lattice simulations of the effective QCD coupling in the MOM scheme, in contrast to that in SF scheme,  suggest that the three flavor system is close to the conformal window. 
In the SF scheme\cite{LNWW92}, the winding number of the system is incorporated through the Chern-Simons actions on the boundary field . It is, however, not clear how the topological winding number of different scale steps are matched .

In the DWF, I define the phase of the quark wave function in the 5th dimension, and adjust the phase so that the wave function on the left wall and that on the right wall are correlated by self-dual instantons. Here, the eigenstate of the triality is considered, but not the winding number of the Chern-Simons action.
 I guess three times more flavor configurations as compared to that of MOM scheme were required in the SF scheme.

In order to justify this conjecture, I first discuss the instanton or self-dual gauge field and their treatment in quaternion basis. Importance of an  instanton in the low energy QCD was pointed out by t'Hooft in 1976\cite{tH76}, and Atiyah and Ward\cite{AW77} showed that it emerges from the minimum action solution for SU(2) Yang-Mills theory. 
They applied the Penrose twistor approach\cite{Pen67} to the instanton problem and showed that there is one to one correspondence between 
a) self-dual solutions of the SU(2) Yang-Mills equation on 4-sphere $S^4$ up to gauge equivalence and 
b) isomorphism classes of 2-dimensional algebraic vector bundles $E$ over projective 3-space $P_3$ satisfying two conditions:

 1) $E$ has a symplectic structure and 

2) The restriction to $E$ of every real line of $P_3$ is algebraically trivial.

The DWF contains the 5th dimension which could play the same role as $z$ in AdS/QCD. It preserves the chiral symmetry in the mass-less limit. We performed a simulation using the DWF gauge configuration of the RBC/UKQCD collaboration\cite{DWF07} by fixing to the Landau gauge and to the Coulomb gauge.

In \cite{SF09}, I analyzed the DWF simulation data using quaternions.
The SU(2) spin algebra is expressible by quaternions ${\bf H}$ invented by Hamilton,  which is a generalization of the complex number $q=w+{\bf i}x+{\bf j}y+{\bf k}z$, endowed with the multiplication rule
\begin{equation}
{\bf i}^2={\bf j}^2={\bf k}^2=-1,\quad
{\bf ij}={\bf k}=-{\bf ji},\quad {\bf jk}={\bf i}=-{\bf kj},\quad {\bf ki}={\bf j}=-{\bf ik}
\end{equation}

The quaternion $q=w+{\bf i}x+{\bf j}y+{\bf k}z$ can be represented by complex $2\times 2$ matrix
\begin{equation}
w+{\bf i}x+{\bf j}y+{\bf k}z=\left(\begin{array}{cc} w+iz & ix+y\\
                                                  ix-y & w-iz\end{array}\right)
\end{equation}

In three dimensional space ${\bf R}^3$ and in seven dimensional space ${\bf R}^7$, the cross product of two basis vectors is expressed by another basis vector.
In the case of ${\bf R}^7$, the rule is given by
\begin{equation}
{\bf e}_i\times {\bf e}_{i+1}={\bf e}_{i+3} 
\end{equation}
where indices are permuted cyclically and translated modulo 7.
The cross product in ${\bf R}^3$ is invariant under rotation of $SO(3)$, but
the cross product in ${\bf R}^7$ is not under rotation of $SO(7)$, but under $G_2$ which is a sub-group of $SO(7)$.  Just as ${\bf a,b,c}$ in ${\bf R}^3$ defined the quaternion algebra ${\bf H}$, ${\bf a,b}$ in ${\bf R}^7$ and the cross product defines an octonion algebra ${\bf O}$.

The content of this paper is as follows. In sect.2, I review the DWF propagator.  I explain, in sect.3 the Cartan's theory of spinor.
The theory of Atiyah-Ward ansatz for the quark propagator is shown in sect.4 and its application in the DWF lattice simulation in MOM scheme is summarized in sect.5.  The implication of triality in lepton flavor phenomenology is given in  sect.6 and on the B-meson leptonic decay etc. are given in sect7. 

\section{DWF lattice simulation}

In the DWF method\cite{Ka92}, the mass of the fermion is
\begin{equation}
m(s)=M\, sign(s)
\end{equation}
The continuum fermion propagator is constructed from 
\begin{equation}
D_F=i\gamma_\mu\partial_\mu+i \gamma_5\partial_s+im(s)
\end{equation}
When a plane wave is assumed in the first four cordinates, the effective one dimensional
problem becomes
\begin{equation}
\hat D_F=i\gamma_5\partial_5+im(s)+\gamma\cdot p
\end{equation}
It is a linear ordinary differential equation of first order and the solution $\psi=U(s,\lambda,p)u(\lambda,p,\alpha)$
where $U(s,\lambda,p)$ is a $4x4$ matrix and $u(\lambda, p,\alpha)$ is an $s$ independent spinor, becomes
\begin{equation}
U(s)=\left\{\begin{array}{ll}\cosh(s \kappa_+)+\kappa_{+}^{-1}\gamma_5[i(\gamma\cdot p-\lambda)-m]\sinh(s\kappa_+)&s\geq 0\\
\cosh(s \kappa_-)+\kappa_{-}^{-1}\gamma_5[i(\gamma\cdot p-\lambda)+m]\sinh(s\kappa_-)&s\leq 0\end{array}
\right.
\end{equation}
where $\kappa_{\pm}=(i\lambda\pm M)^2+p^2$.

The eigenvalue equation on the lattice in the $s$ space becomes
\begin{eqnarray}
&&\frac{1}{2}\gamma_5[U(s+1)-U(s-1)]+\frac{1}{2}[U(s+1)+U(s-1)-2U(s)]\nonumber\\
&&+\left(i\sum_{\mu=1}^{4}\gamma_\mu \sin(p_\mu)-F(p)+m(s)+i\lambda\right)U(s)=0
\end{eqnarray}
where $F(p)=\sum_{\mu=1}^4[1-\cos(p_\mu)]$.

The positive chirality component $U_+(s)$ and the negative chirality component
$U_-(s)$ satisfy the relations
\begin{equation}
U_+(s+1)=[1+F(p)-i\lambda-m(s)]U_+(s)-i\sum_{\mu=1}^4\sigma_\mu\sin(p_\mu)U_-(s)
\end{equation}
\begin{equation}
U_-(s-1)=[1+F(p)-i\lambda-m(s)]U_-(s)-i\sum_{\mu=1}^4\sigma_\mu\sin(p_\mu)U_-(s)
\end{equation}
\begin{eqnarray}
U(s+1)&=&T^+ U(s), \qquad s>0\\
U(s-1)&=&T^- U(s), \qquad s\leq 0
\end{eqnarray}
\begin{equation}
T^+=\left(\begin{array}{cc}1+F(p)-i\lambda-M&-i\sum_{\mu=1}^4\sigma_\mu\sin(p_\mu)\\
i\sum_{\mu=1}^4\bar\sigma_\mu \sin(p_\mu)& \frac{1+G(p)}{1+F(p)-i\lambda-M}\end{array}
\right)
\end{equation}
where $G(p)=\sum_{\mu=1}^4 \sin^2(p_\mu)$

\begin{equation}
{\mathcal G}_F=D^\dagger_F {\mathcal G}^{chiral}(s,s';p)=\frac{1+\gamma_5}{2}\frac{e^{(-M|s|-M|s'|)}}{\gamma\cdot p}
\end{equation}

This expression shows infrared singularity.  This singularity comes from the bound state
pole in the lefthanded quark propagator.
\begin{equation}
G_L(p)=\frac{1}{\sum_{\mu=1}^4 \bar p_\mu^2+M^\dagger(p)M(p)}, \quad G_R(p)=\frac{1}{\sum_{\mu=1}^4 \bar p_\mu^2+M(p) M^\dagger(p)}, 
\end{equation}
When $M>0$ and $s\geq 0\geq s'$ 
\begin{equation}
[G_L(p)]_{ss'}=A_L(p)e^{-\alpha_+(p)s+\alpha_-(p)s'} 
\end{equation}
has the singularity, which is however cancelled by the gauge part\cite{NaNe93,Sha93}.

\begin{figure}
\begin{minipage}[b]{0.47\linewidth}
\begin{center}
\includegraphics[width=6cm,angle=0,clip]{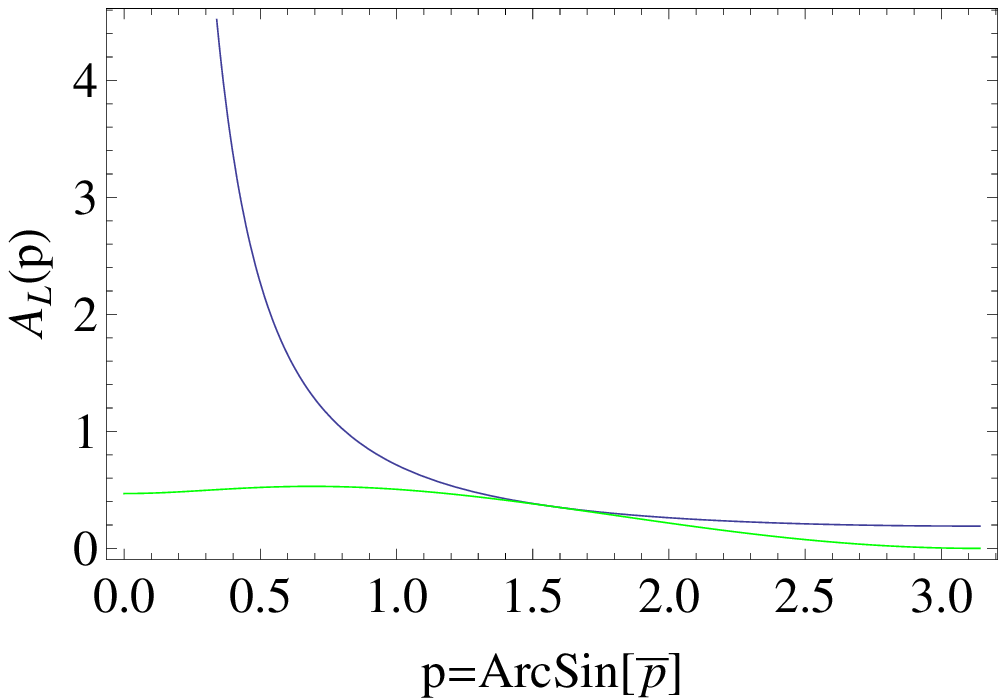}
\caption{Left handed chiral fermion amplitude $A_L(p)$(blue) and $A_L(p)\bar p^2$(green).  $M=1$  }
\label{AL}
\end{center}
\end{minipage}
\hfill        
\begin{minipage}[b]{0.47\linewidth}
\begin{center}
\includegraphics[width=6cm,angle=0,clip]{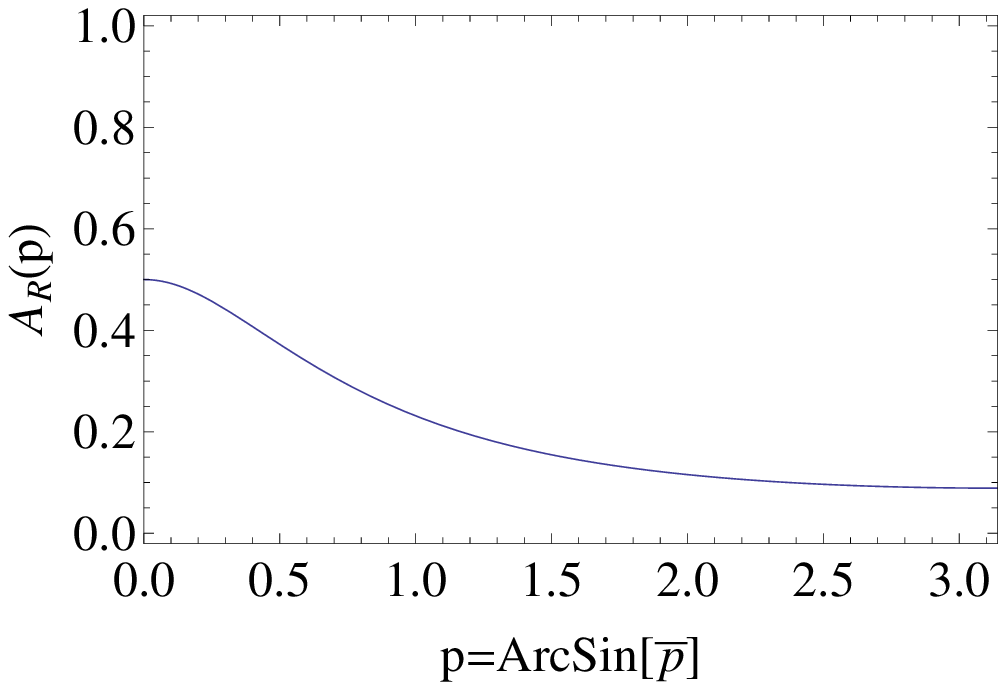}
\caption{Right handed chiral fermion amplitude $A_R(p)$. $M=1$ }
\label{AR}
\end{center}
\end{minipage}
\end{figure}

The kinetic energy extracted from the propagator at $s=s'=0$ becomes
\begin{equation}
\frac{1}{A_L(p)\bar p^2}i\sum_{\mu=1}^4[\gamma_\mu\bar p_\mu]P_R
\end{equation} 
where $A_L(p)\bar p^2$ can be interpreted as the normalization factor of the zero mode.
It is finite at $p=0$ as shown in Fig.\ref{AL}. Here I took $M=1$.

When the domain wall height $M$ is taken to be negative $M=-1.8$ as assumed in\cite{DWF07}, the amplitude $A_R(p)$becomes singular in the infrared. 
\begin{figure}
\begin{minipage}[b]{0.47\linewidth}
\begin{center}
\includegraphics[width=6cm,angle=0,clip]{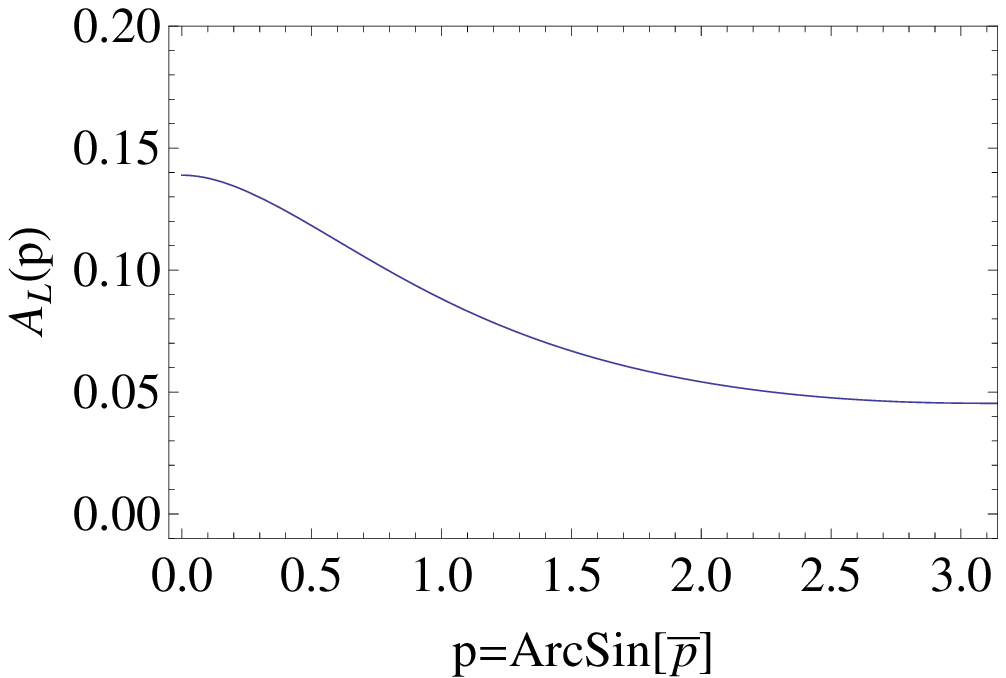}
\caption{Negative mass chiral fermion amplitude  $A_R(p)$ and $A_R(p)\bar p^2$(green).$M=-1.8$  }
\label{AL1}
\end{center}
\end{minipage}
\hfill        
\begin{minipage}[b]{0.47\linewidth}
\begin{center}
\includegraphics[width=6cm,angle=0,clip]{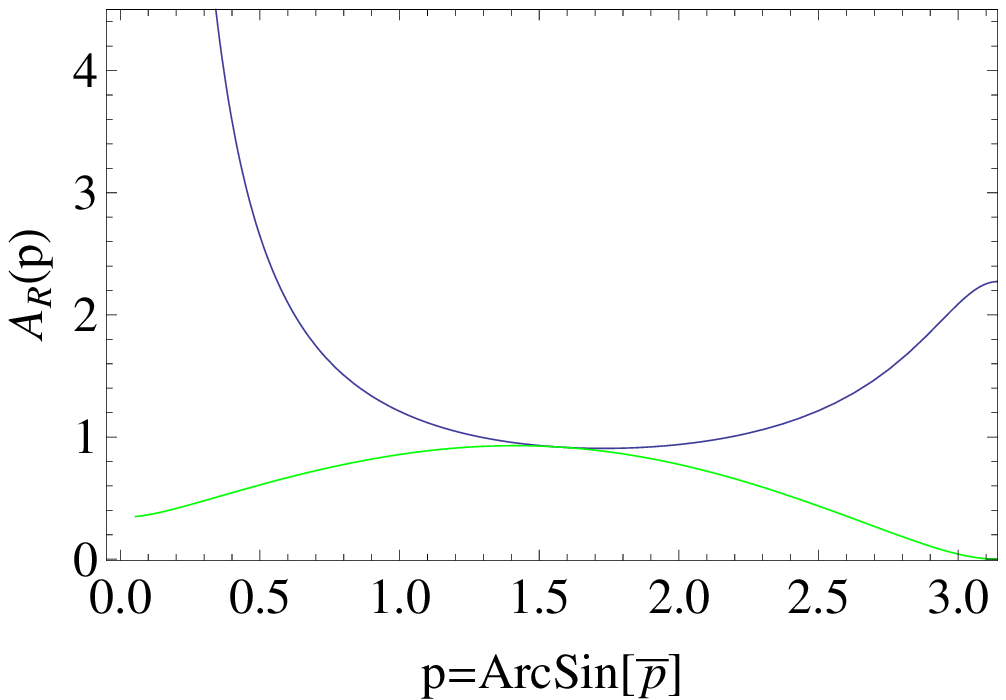}
\caption{Negative mass chiral fermion amplitude $A_L(p)$(blue).  $M=-1.8$ }
\label{AR1}
\end{center}
\end{minipage}
\end{figure}

\section{Cartan's theory of spinors}

The quaternion space spanned by $\bf H$ spanned by $\{1,i,j,k\}$ can be extended by introducing
a new imaginary unit $\l$ and make the octonion spsce ${\bf O}={\bf H}+{\bf H}l$ spanned by $\{1,i,j,k,l,il,ij,ik\}$.

Cartan studied $2\nu$ dimensional Euclidean space $E_{2\nu}$ and semi-spinor $\xi_\alpha$ with components of odd or even number of indices are zero.
He defined $\phi$ as those of even number of indices are 0 and $\psi$ as those of odd number of indices are 0, and arbitrary $\nu-$vector $X_{(\nu)}$.
Triality transformation rotates 24 dimensional bases defined by Cartan\cite{Cartan66}.

\begin{eqnarray}
\{\xi_0, \xi_1, \xi_2, \xi_3, \xi_4\},\quad
\{\xi_{12}, \xi_{31}, \xi_{23}, \xi_{14}, \xi_{24}, \xi_{34}\},\quad
\{\xi_{123}, \xi_{124}, \xi_{314}, \xi_{234}, \xi_{1234}\}
\nonumber\\
\{x^1, x^2, x^3, x^4\}, \quad \{x^{1'}, x^{2'}, x^{3'}, x^{4'}\}
\end{eqnarray}

The trilinear form in these bases is
\begin{eqnarray}
{\mathcal F}&&=^t\phi CX\psi=x^1(\xi_{12}\xi_{314}-\xi_{31}\xi_{124}-\xi_{14}\xi_{123}+\xi_{1234}\xi_1)\nonumber\\
&&+x^2(\xi_{23}\xi_{124}-\xi_{12}\xi_{234}-\xi_{24}\xi_{123}+\xi_{1234}\xi_2)\nonumber\\
&&+x^3(\xi_{31}\xi_{234}-\xi_{23}\xi_{314}-\xi_{34}\xi_{123}+\xi_{1234}\xi_3)\nonumber\\
&&+x^4(-\xi_{14}\xi_{234}-\xi_{24}\xi_{314}-\xi_{34}\xi_{124}+\xi_{1234}\xi_4)\nonumber\\
&&+x^{1'}(-\xi_{0}\xi_{234}+\xi_{23}\xi_{4}-\xi_{24}\xi_{3}+\xi_{34}\xi_2)\nonumber\\
&&+x^{2'}(-\xi_{0}\xi_{314}+\xi_{31}\xi_{4}-\xi_{34}\xi_{1}+\xi_{14}\xi_3)\nonumber\\
&&+x^{3'}(-\xi_{0}\xi_{124}+\xi_{12}\xi_{4}-\xi_{14}\xi_{2}+\xi_{24}\xi_1)\nonumber\\
&&+x^{4'}(\xi_{0}\xi_{123}-\xi_{23}\xi_{1}-\xi_{31}\xi_{2}-\xi_{12}\xi_3)
\end{eqnarray}
There are three semi-spinors which have a quadratic form which is invariant with respect to the group of rotation

\begin{eqnarray}
&&\Phi={^t\phi}C\phi=\xi_0\xi_{1234}-\xi_{23}\xi_{14}-\xi_{31}\xi_{24}-\xi_{12}\xi_{34}\nonumber\\
&&\Psi={^t\psi}C\psi=-\xi_1\xi_{234}-\xi_{2}\xi_{314}-\xi_{3}\xi_{124}+\xi_{4}\xi_{123}
\end{eqnarray}
and the vector
\begin{equation}
F=x^1 x^{1'}+x^2 x^{2'}+x^3 x^{3'}+x^4 x^{4'}
\end{equation}

The trilinear form $\mathcal F$ is invariant and the three quadratic forms interchange, under transformations which are classified as $G_{(23)}, G_{(12)}$ and $G_{(13)}$,
which are shown in the Appendix . The product $^t(G_{(12)})G_{(13)}$ generates $G_{(132)}$ and $^t(G_{(13)})G_{(12)}$ generates $G_{(123)}$. \footnote{The definition in \cite{Cartan66}, $G_{(12)}G_{(13)}=G_{(132)}$ and   $G_{(13)}G_{(12)}=G_{(123)}$ are misprints.}

The semi-spinor $\xi_h$, $\xi_{ij}$, $\xi_{klm}$ and $\xi_{1234}$ are devided into two classes. The first type satisfies
\begin{equation}
\Phi=\xi_0\xi_{1234}-\xi_{12}\xi_{34}-\xi_{23}\xi_{14}-\xi_{31}\xi_{24}=0
\end{equation}
and the second type satisfies
\begin{equation}
\Psi=-\xi_1\xi_{234}-\xi_{2}\xi_{134}+\xi_{3}\xi_{124}+\xi_{4}\xi_{123}=0
\end{equation}

With a use of quaternion bases $1,{\Vec i},{\Vec j}, {\Vec k}$, the spinors $\phi$ and $C\phi=\psi'$ are defined as
\begin{equation}
\phi=\xi_0+\xi_{14}{\Vec i}+\xi_{24}{\Vec j}+\xi_{34}{\Vec k}
=\left(\begin{array}{cc}\xi_0+i\xi_{34},&i\xi_{14}-\xi_{24}\\
                                   i\xi_{14}+\xi_{24}&\xi_0-i\xi_{34}\end{array}\right)
\end{equation}
\begin{equation}
C\phi=\xi_{1234}-\xi_{23}{\Vec i}-\xi_{31}{\Vec j}-\xi_{12}{\Vec k} 
=\left(\begin{array}{cc}\xi_{1234}-i\xi_{12},&-i\xi_{23}+\xi_{31}\\
                                  -i\xi_{23}-\xi_{31}&\xi_{1234}+i\xi_{12}
\end{array}\right)
\end{equation}
Similarly, $\psi$ and $C\psi=\psi'$ are defined as
\begin{equation}
\psi=\xi_4+\xi_1{\Vec i}+\xi_2{\Vec j}+\xi_3{\Vec k}
=\left(\begin{array}{cc}\xi_4+i\xi_{3},&i\xi_{1}-\xi_{2}\\
                                   i\xi_{1}+\xi_{2}&\xi_4-i\xi_{3}\end{array}\right)
\end{equation}
\begin{equation}
C\psi=\xi_{123}-\xi_{234}{\Vec i}-\xi_{314}{\Vec j}-\xi_{124}{\Vec k}
=\left(\begin{array}{cc}\xi_{123}-i\xi_{124},&-i\xi_{234}+\xi_{314}\\
                                  -i\xi_{234}-\xi_{314}&\xi_{123}+i\xi_{124}\end{array}\right)
\end{equation}
When two spinors $\psi$ and $C\psi$ or $\phi$ and $C\phi$ are combined together, they are expressed by octonion, and the automorphism in the space of octonion is $G_2$ group, which is exceptional Lie group with 14 dimensional representation. In $G_2$, there is a specific automorphism, which is called triality. In the following subsection, I study the structure of the $G_2$ group.

The trilinear form ${^t\phi} CX\psi$ which is invariant under triality transformation contains the term $x^1\xi_{14}\xi_{123}$, $x^2\xi_{24}\xi_{123}$ and $x^3\xi_{34}\xi_{123}$ which transforms by applying $G_{132}$ to $-\xi_{234}x^{1'}\xi_0$, $-\xi_{314}x^{2'}\xi_0$ and $-\xi_{124}x^{3'}\xi_0$, respectively.

The vector $x^i$ and $x^{i'}$ are Pl\"ucker coordinates of the line which joins the two points $\phi$ and $\psi'$ of quark specified by $\phi$ and quark specified by $\psi'$.

\section{The Atiyah-Ward ansatz in the quark propagator}
The null twister $\xi$ defined by Penrose is a combination of two spinors $\lambda=(\lambda_{\dot{a}})$
and $\mu=(\mu^a)$ in the complex projective space ${\bf CP}^3$ space.  It satisfies
\begin{eqnarray}
&&(\mu^{0*}, \mu^{1*},\lambda_{\dot{0}}^*, \lambda_{\dot{1}}^*)\left(\begin{array}{cccc} 0&0&1&0\\
                                                                           0&0&0&1\\
                                                                           1&0&0&0\\
                                                                           0&1&0&0\end{array}\right)
\left(\begin{array}{c} \mu^0\\
                          \mu^1\\
                           \lambda_{\dot{0}}\\
                        \lambda_{\dot{1}}\end{array}\right)  \nonumber\\
&&=\bar\lambda_a\mu^a+\bar\mu^{\dot{a}}\lambda_{\dot{a}}=0    
\end{eqnarray}

The components of the null twister  satisfies
\begin{equation}
\mu^a=x^{a\dot{a}}\lambda_{\dot{a}}
\end{equation}

An open subset of the twister space ${\bf PT}$ defined as 
${\bf PT}'=\{[\xi]\in {\bf PT}
|\xi=(\mu,\lambda),\lambda=(\lambda_{\dot{0}},\lambda_{\dot{1}})\ne (0,0)\}$
allows a mapping on $\bf E$
\begin{equation}
\left(\begin{array}{cc}y&-\bar z\\
                        z&\bar y\end{array}\right)=\left(\begin{array}{cc}\mu^0&-\bar{\mu^1}\\
                                                                       \mu^1&\bar{\mu^0}\end{array}\right)
\left(\begin{array}{cc}\mu^0&-\bar{\lambda^1}\\
                                                 \lambda^1&\bar{\lambda^0}\end{array}\right)^{-1}
\end{equation}

The Atiyah-Ward ansatz for the quark propagator was adopted by Corrigan and Goddard\cite{CG81} in
the transformation function $g(\omega,\pi)$ where $\omega$ and $\pi$ are complex two-spinor which satisfy in the ${\bf CP}^3$ space,
\[
g(\lambda\omega,\lambda\pi)=g(\omega,\pi), \quad det\, g=1.
\]

When  $x=x^0-i {\Vec x}\cdot{\Vec\sigma}$ is a quaternion and $\omega=x\pi$, $g(\omega,\pi)$ can be expressed as
\[
g(x\pi,\pi)=h(x,\zeta)k(x,\zeta)^{-1}
\]
where $\displaystyle\zeta=\frac{\pi_1}{\pi_2}$, $h(x,\zeta)$ is regular in $|\zeta|>1-\epsilon$ and $k(x,\zeta)$ is regular in $|\zeta|<1+\epsilon$.

In \cite{CG81},  the Ansatz
\begin{eqnarray}
g_0&&=\left(\begin{array}{cc}e^{-\nu}&0\\
                           0&e^\nu\end{array}\right)
\left(\begin{array}{cc}\zeta^1&\rho\\
                       0&\zeta^{-1}\end{array}\right)
\left(\begin{array}{cc}e^{\mu}&0\\
                           0&e^{-\mu}\end{array}\right)\nonumber\\
&&=\left(\begin{array}{cc}e^\gamma\zeta^1&f(\gamma,\zeta)\\
                       0&e^{-\gamma}\zeta^{-1}\end{array}\right)
\end{eqnarray}
was proposed as the transformation matrix. Here $\gamma=\mu-\nu$. 

In \cite{Tak05}, a solution of equation $\Psi_\infty g=\Psi_0$, where
\begin{equation}
\Psi_0=\left(\begin{array}{cc} \phi_{11}&\phi_{12}\\
                                \phi_{21}&\phi_{22}\end{array}\right),
\qquad
\Psi_\infty=\left(\begin{array}{cc} \psi_{11}&\psi_{12}\\
                                \psi_{21}&\psi_{22}\end{array}\right),
\end{equation}
under the boundary condition $\Psi(x,\infty)=1$ was considered. In our system, I identify
$\Psi_0$ and $\Psi_\infty$ as gauge configuration on the left domain wall and the right domain wall.

\section{The quark propagator in Coulomb gauge and the QCD effective coupling}

I first perform a gauge fixing in Landau gauge using the over relaxation method and then fix to the Coulomb gauge without touching the $A_4$ component\cite{FN07}. I then measure the quark propagator and fix the gauge in the space of the 5th dimension between the domain walls.

 Since we adopt the representation of fermion spinors such that the $\gamma_5$ is diagonal, the fermion field $\psi$ can be gauge transformed in the fifth dimension as
\begin{equation}
\psi\to e^{i\eta\gamma_5}\psi, \qquad \bar\psi \to \bar\psi e^{-i\eta\gamma 5},
\end{equation}
such that on the left and on the right domain wall, the propagator becomes approximately real by choosing \cite{SF09}
\begin{equation}
|e^{i\theta L}e^{i\eta}-1|^2+|e^{i\theta R}e^{-i\eta}-1|^2
\end{equation}
be minimum.

The equation has two sets of solutions, which can be obtained numerically. I assign  
\begin{equation}
\Psi_{0, L/R}=
\left(\begin{array}{cc}a_{0}& b_{0}\\
                       c_{0}& d_{0}\end{array}\right)_{L/R} {\rm and }\quad
\Psi_{L_s-1,L/R}=\left(\begin{array}{cc}a_{L_s-1}& b_{L_s-1}\\
                       c_{L_s-1}& d_{L_s-1}\end{array}\right)_{L/R}
\end{equation}
from the color diagonal components of the lattice data.

Using the lattice wave function $\widetilde{\psi(p,k)}$ and the plane wave $\chi(p,k)$, where $k$ is the coordinate in the 5th dimension, I define
\begin{eqnarray}
&&Tr\langle \chi(p_x,0)\widetilde{\psi_{L/R}(p_x,0)}\rangle\sigma_1+
Tr\langle\chi(p_y,0)\widetilde{\psi_{L/R}(p_y,0)}\rangle\sigma_2+\nonumber\\
&&Tr\langle \chi(p_z,0)\widetilde{\psi_{L/R}(p_z,0)}\rangle\sigma_3+
Tr\langle\chi(p_t,0)\widetilde{\psi_{L/R}(p_t,0)}\rangle iI\nonumber\\
&&\propto a_x^0\sigma_1+a_y^0 \sigma_2+a_z^0 \sigma_3+a_t^0 iI
\end{eqnarray} 
and
\begin{eqnarray}
&&Tr\langle \chi(p_x,L_s-1)\widetilde{\psi_{L/R}(p_x,L_s-1)}\rangle\sigma_1+
Tr\langle\chi(p_y,L_s-1)\widetilde{\psi_{L/R}(p_y,L_s-1)}\rangle\sigma_2+\nonumber\\
&&Tr\langle \chi(p_z,L_s-1)\widetilde{\psi_{L/R}(p_z,L_s-1)}\rangle\sigma_3+
Tr\langle\chi(p_t,L_s-1)\widetilde{\psi_{L/R}(p_t,L_s-1)}\rangle iI\nonumber\\
&&\propto a_x^{L_s-1}\sigma_1+a_y^{L_s-1} \sigma_2+a_z^{L_s-1} \sigma_3+a_t^{L_s-1} iI
\end{eqnarray}

If the configulation on the left domain wall and the right domain wall are correlated by self dual gauge field,  the transformation matrix $g(\gamma,\zeta)$ gives a relation,
\begin{equation}
\left(\begin{array}{cc}a_{L_s-1}& b_{L_s-1}\\
                       c_{L_s-1}& d_{L_s-1}\end{array}\right)
\left(\begin{array}{cc}\zeta^{1}e^\gamma& f\\
                       0&\zeta^{-1}e^{-\gamma}\end{array}\right)
=\left(\begin{array}{cc}\zeta^{1}e^{-\gamma}& \bar f\\
                       0&\zeta^{-1}e^{\gamma}\end{array}\right)
\left(\begin{array}{cc}a_{0}& b_{0}\\
                       c_{0}& d_{0}\end{array}\right)
\end{equation}
where $\displaystyle \bar f=\overline{f(\bar\gamma,-\frac{1}{\bar\zeta})}$.
 
In our 5-dimesional domain wall fermion case,  the introduced phase in the 5th direction $i\eta$ is incorporated in $\mu$ and  $-i\eta$ is incorporated in $\nu$. 

The function $f$ and $\bar f$ taken in \cite{CG81} is
\begin{equation}
f=\frac{d_0 e^\gamma-\frac{1}{a_{L_s-1}}e^{-\gamma}}{\psi},\quad 
\bar f=\frac{\frac{1}{d_{L_s-1}} e^\gamma-a_0e^{-\gamma}}{\psi}.
\end{equation}

In general $c_0$ and $c_{L_s-1}$ are polynomials of $\zeta$.
 I define
\begin{equation}
\psi=\hat c_{-1}\zeta^{-1}+\hat c_{1}\zeta^{1}+\delta
\end{equation}
where $\hat c_{-1}=c_{L_s-1}$ and $\hat c_{1}=c_0$ and $\delta$ is a constant. They are defined by minimizing the difference
\begin{equation}
\Delta L/R=\left(\begin{array}{cc}a_{L_s-1}& b_{L_s-1}\\
                       c_{L_s-1}& d_{L_s-1}\end{array}\right)_{L/R}
\left(\begin{array}{cc}\zeta^{1}e^\gamma& f\\
                       0&\zeta^{-1}e^{-\gamma}\end{array}\right)
-\left(\begin{array}{cc}\zeta^{1}e^{-\gamma}& \bar f\\
                       0&\zeta^{-1}e^{\gamma}\end{array}\right)
\left(\begin{array}{cc}a_{0}& b_{0}\\
                       c_{0}& d_{0}\end{array}\right)_{L/R}.
\end{equation}
It is the condition for the left wall and the right wall are correlated by the self-dual gauge transformation.

Since $\zeta$ is a parameter in ${\bf CP}^3$ space, I adopt an ansatz $\zeta^1 e^\gamma=
\sqrt{\frac{c_0}{c_{L_s-1}}\frac{d_{L_s-1}}{d_0}}e^{i\eta}$, 
and derived $e^{i\eta}$ and the $\delta$ by solving the simultaneous equation (\ref{gamma}) using Mathematica\cite{wolfram}
\begin{eqnarray}
\left\{\begin{array}{l}
a_{L_s-1}f+b_{L_s-1}\zeta^{-1}e^{-i\eta}=b_0\zeta^{-1}e^{-i\eta}+d_0 \bar f\nonumber\\
c_{L_s-1}f+d_{L_s-1}\zeta^{-1}e^{-i\eta}=d_0\zeta^{-1}e^{i\eta}.
\end{array}
\right.
\end{eqnarray}\label{gamma}

From the numerical practice, I observe that the deviation $\Delta L/R$ becomes small when $|\delta|$ is large, i.e. when $\psi\sim\delta\sim \frac{1}{f}$ is large. 
When $\delta$ is large
\begin{equation}
\frac{1}{\psi}=\frac{1}{ {c_{-1}}{\zeta^{-1}} +\delta+c_1\zeta}=\frac{1}{\delta}-({c_{-1}}{\zeta^{-1}}+c_1\zeta)\frac{1}{\delta}
+\frac{1}{2}({c_{-1}}{\zeta^{-1}}+c_1\zeta)^2\frac{1}{\delta^2}
+\cdots
\end{equation}
and the Laurent series of $f$ converges fast.

In the form factor calculation I use the overlap of gauge transformed $\chi$ and the plane wave at the middle of the domain walls.

Corresponding to the sample-wise selection of larger absolute value of $\delta$, i.e. a large zero-mode component, I assign a parameter $ind=1$ for larger and 2 for smaller, and multiply the phase $(-1)^{ind}$ to the expectation value of the quark wave function that is used in the calculation of the effective mass of the quark. 
This phase factor can be absorbed in the phase $e^{i\eta}$ introduced to approximately satisfy the U(1) real condition.

\section{Lepton flavor phenomenology}
The mass of a fermion is produced by Yukawa term which couples the boundary layer $s=N$
of a quark of a chilarity and the boundary layer $s=0$ of a quark of an opposite chilarity. 
In the minimal supersymmetric standard model(MSSM), the left-handed quark is defined as
\begin{equation}\Psi_{Q_L}=\left(\begin{array}{c}\Psi_{u_L}\\
                                                             \Psi_{d_L}\end{array}\right)
\end{equation}
which have $SU(3)_c\times SU(2)_L\times U(1)_Y$ quantum numbers ${\bf 3},{\bf 2},1/3$.
The right-handed quarks $u_R$ and $d_R$ are $SU(2)_L $ singlet and they have quantum numberd
${\bf 3},{\bf 2},{4/3}$ and ${\bf 3},{\bf 2},{-2/3}$, respectively.

In the MSSM, the Higgs field is defined as
\begin{equation}
H=\left(\begin{array}{c} H^+\\
                               H^0\end{array}\right)
\end{equation}
and 
\begin{equation}
H^{\dagger T}=\left(\begin{array}{c}(H^+)^\dagger\\
                                             (H^0)\dagger\end{array}\right)
=\left(\begin{array}{c}H^-\\
                            H^{0\dagger}\end{array}\right)                                                                   
\end{equation}
The vacuum expectation value of $H$ is
\begin{equation}
\langle H^\dagger H\rangle=\nu^2
\end{equation}

The down quark mass is produced by the Yukawa coupling
\begin{equation}
y_d\bar\Psi_{Q_L}H \Psi_{d_R}+h.c.
\end{equation}
and the up quark mass is produced by
\begin{equation}
y_u\Psi_{Q_L}i\tau_2 H^{\dagger T}\Psi_{u_R}+h.c.
\end{equation}

The masses of $u,c,t$ quarks $m_{u,c,t}$ and $d,s,b$ quarks $m_{d,s,b}$ are expressed as 
\begin{equation}
m_{u,c,t}=\nu_u y_{u,c,t},  \qquad m_{d,s,b}=\nu_d y_{d,s,b}
\end{equation}

When the spinors $\phi$ and $\psi$ are assigned to a lepton, $e, \mu$ or $\tau$, the electro-magnetic interaction requires they belong to the same family $G_{(23)}$, $G_{(12)}$ or $G_{(13)}$. 
The  quark interaction with neutrino is expected to be blind to the lepton flavor, and the neutrino mixing occurs.

Leptons are characterized by the flavor triality. Non-Abelian discrete flavor symmetry $A_4$ is assumed to be broken to $Z_3$ in the quarks and by assigning $\omega=e^{2\pi i/3}$ the Left-handed quarks and leptons are classified \cite{Ma10,BBJ81}
\begin{eqnarray}
\qquad\begin{array}{ccc}
 u,d,e\sim 1 &\qquad\qquad  c,s,\mu\sim \omega^2 &\qquad\qquad t,b,\tau\sim \omega
\end{array}\nonumber\\
\qquad\left(\begin{array}{cccc}
 \nu_e & u_r & u_g & u_b\\
 e^- & d_r & d_g & d_b\end{array}\right)_L
\quad
\left(\begin{array}{cccc}
 \nu_\mu & c_r & c_g & c_b\\
 \mu^- & s_r & s_g & s_b\end{array}\right)_L
\quad
\left(\begin{array}{cccc}
 \nu_\tau & t_r & t_g & t_b\\
 \tau^- & b_r & b_g & b_b\end{array}\right)_L
\nonumber
\end{eqnarray}
\vskip 0.3 true cm
and the Right-handed quarks and leptons
\begin{eqnarray}
\qquad\begin{array}{ccc}
 u^c, d^c, e^c\sim 1 &\qquad\qquad  c^c,s^c,\mu^c\sim \omega &\qquad\qquad t^c,b^c,\tau^c\sim \omega^2.
\end{array}\nonumber\\
\qquad\left( \begin{array}{cccc}
e^- & u_r & u_g & u_b\\
           & d_r & d_g & d_b\end{array}\right)_R
\quad
\left( \begin{array}{cccc}
\mu^- & c_r & c_g & c_b\\
           & s_r & s_g & s_b\end{array}\right)_R
\quad
\left( \begin{array}{cccc}
 \tau^- & t_r & t_g & t_b\\
           & b_r & b_g & b_b\end{array}\right)_R
\nonumber
\end{eqnarray}

\vskip 0.3 true cm
There are self-dual gauge vector fields in the color-isospin space and in color-spin space.  The vector field has the $SU(5)\supset SU(3)_C\times SU(2)_W\times U(1)$ symmetry.  In the standard model, $W,Z$ and electromagnetic field $A^{em}$ are unified in Weinberg-Salam model. The gluon fields are denoted $A_1,A_2,A_3$ and $A_0$.

\[
\left( \begin{array}{cccc}
  W^+&  W^- &  Z&  A^{em}\\
  A_1& A_2 &  A_3&  A_0\end{array}\right)_{t1} \quad
\left (\begin{array}{cccc}
  W^+&  W^-&   Z&  A^{em}\\
 A_1& A_2 & A_3&  A_0\end{array}\right)_{t2}\quad 
\left (\begin{array}{cccc}
  W^+&  W^-&  Z&  A^{em}\\
 A_1& A_2 & A_3&  A_0\end{array}\right)_{t3}
\]

In the following figures, I assign the self dual gauge fields $W$ or $A$ as
\[
\left (\begin{array}{cccc}
  x_1'& x_2'& x_3'& x_4'\\
  x_1&   x_2 & x_3& x_4\end{array}\right)
\]
which has their specific triality.

 The quarks ${q}=u,d,s,b,c,t$ has $G_2\supset SU(3)_C\times SU(2)_{spin}$ symmetry, but triality blind. 

Since vector fields are defined from spinors as Pl\"ucker coordinates, the gluons that interact with quarks but not with $e,\mu$ and $\tau$,  the triality symmetry for gauge bosons $A^{em}, W$ and $Z$  and gluons could be qualitatively different.
In the case of a color-isospin triality, electron number, muon number and tau meson number are well defined, but $u,d,s,c,b$ and $t$ quarks mix with each other. I expect gluon-weak current coupling preserves flavor triality and color-spin triality, but the quark-gluon coupling is blind to flavor triality\cite{Ma10} and to color-spin triality. The absence of flavor changing neutral current could be due to the triality selection, although the GIM mechanism\cite{GIM70} could also play a role.

The lattice simulation is the most efficient method to clarify the mechanism of confinement and chiral symmetry breaking, since they are essentially non-perturbative phenomena.

In astrophysics, most astronomical data are taken from electro-magnetic detectors, and electro-magnetic wave from a triality family that is different from that of the electro-magnetic probe would evade the detection.

\section{Discussion}
I studied possible roles of triality in color-spin space and its consequence in the IR-QCD, and extension of flavor-preserving self-dual gauge field to flavor changing gauge boson. 

It is not evident that the bispinor system proposed by Cartan actually explains the lattice simulation results.
 However, the difference of a critical flavor number in SF scheme, which is three times larger than that in the MOM scheme could be understood, if the boundary condition imposed in SF scheme, in which at each recursive renormalization step, matching large L lattice data to small L lattice data is performed, requires more freedom in the functional space of fermions.

From the symmetry of octonions, Cartan constructed the vector fields $x_1,x_2,x_3, x_4$ and their duals $x_1',x_2',x_3',x_4'$ from spinors. Since I consider a self-dual vector field that couple to spinors, I use $x_1,x_2,x_3$ for the self-dual gluon fields or $W^{\pm}$ fields. 
From the invariant trilinear form, I consider the Coulomb potential and the transverse component. I omit diagrams in which $x_4$ is exchanged, since in magneto static QCD\cite{IKRV06}, one can choose $A_0=0$, and ignore  $x_4$ exchange in self-energy diagrams.
\begin{figure}
\begin{minipage}[b]{0.47\linewidth}
\begin{center}
\includegraphics[width=6cm,angle=0,clip]{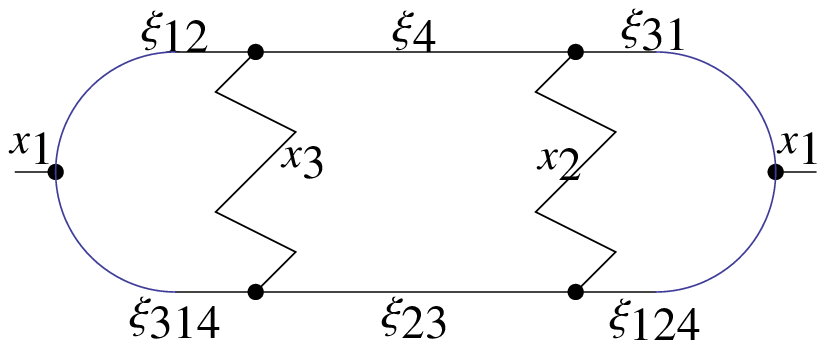}
\caption{Transverse gluon or $W^{\pm}$ self energy diagram: g611a}
\label{11a}
\end{center}
\end{minipage}
\hfill
\begin{minipage}[b]{0.47\linewidth}
\begin{center}
\includegraphics[width=6cm,angle=0,clip]{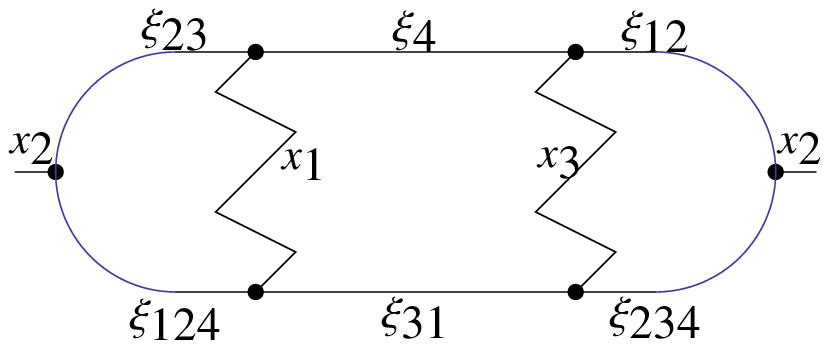}
\caption{Transverse gluon or $W^{\pm}$ self energy diagram: g622a}
\label{22a}
\end{center}
\end{minipage}
\end{figure}

\begin{figure}
\begin{minipage}[b]{0.47\linewidth}
\begin{center}
\includegraphics[width=6cm,angle=0,clip]{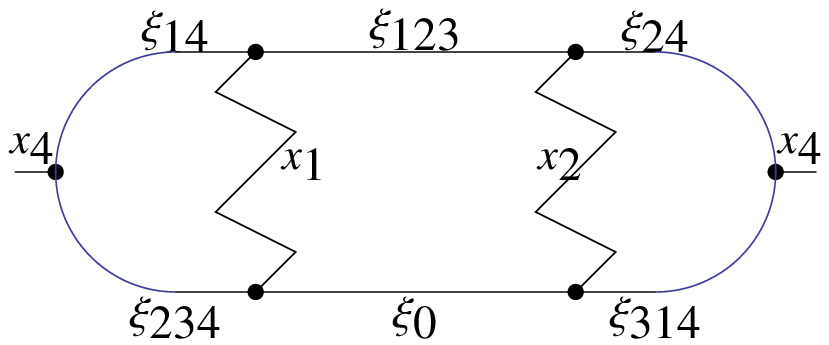}
\caption{Coulomb potential self energy diagram: g644a}
\label{44a}
\end{center}
\end{minipage}
\hfill
\begin{minipage}[b]{0.47\linewidth}
\begin{center}
\includegraphics[width=6cm,angle=0,clip]{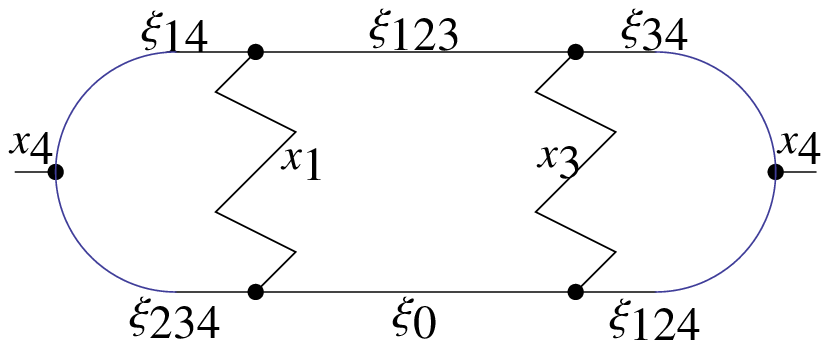}
\caption{Coulomb potential self energy diagram: g644b}
\label{44b}
\end{center}
\end{minipage}
\end{figure}

\begin{figure}
\begin{center}
\includegraphics[width=6cm,angle=0,clip]{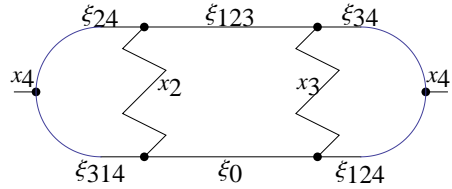}
\caption{Coulomb potential self energy diagram: g644c}
\label{44c}
\end{center}
\end{figure}

The three loop self energy could play an important role in magnetic mass problem\cite{Linde80} in finite temperature QCD.

\begin{figure}
\begin{minipage}[b]{0.47\linewidth}
\begin{center}
\includegraphics[width=5cm,angle=0,clip]{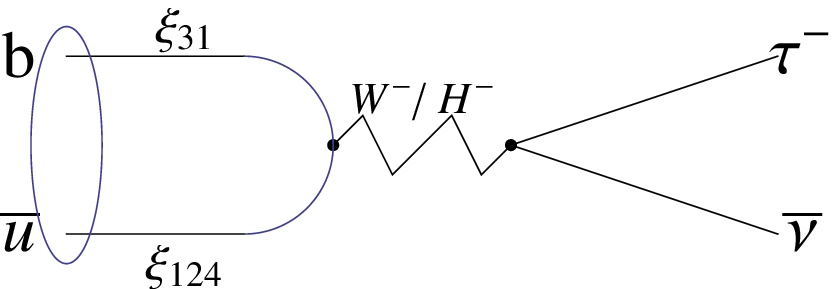}
\caption{The decay of $\bar B$ to $\tau^- \nu$. Transverse polarization tree diagram 1a.}
\label{Bdecay1a}
\end{center}
\end{minipage}
\hfill
\begin{minipage}[b]{0.47\linewidth}
\begin{center}
\includegraphics[width=5cm,angle=0,clip]{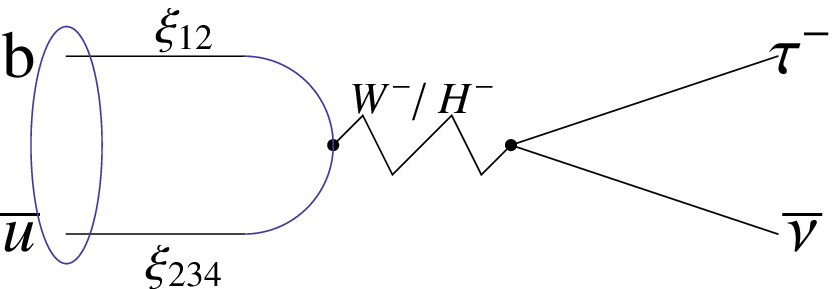}
\caption{Transverse polarization tree diagram 2a.}
\label{Bdecay2a}
\end{center}
\end{minipage}
\end{figure}

\begin{figure}
\begin{minipage}[b]{0.47\linewidth}
\begin{center}
\includegraphics[width=5cm,angle=0,clip]{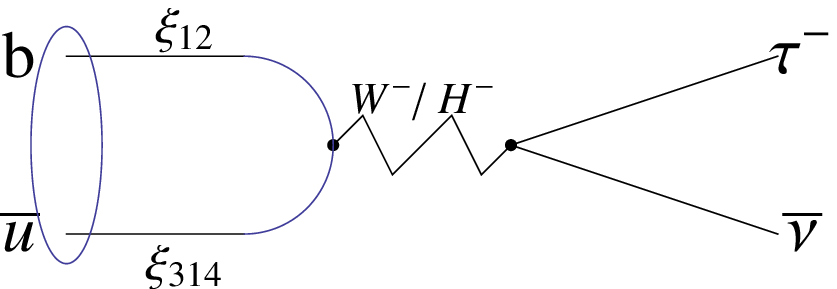}
\caption{Transverse polarization tree diagram 1b.}
\label{Bdecay1b}
\end{center}
\end{minipage}
\hfill
\begin{minipage}[b]{0.47\linewidth}
\begin{center}
\includegraphics[width=5cm,angle=0,clip]{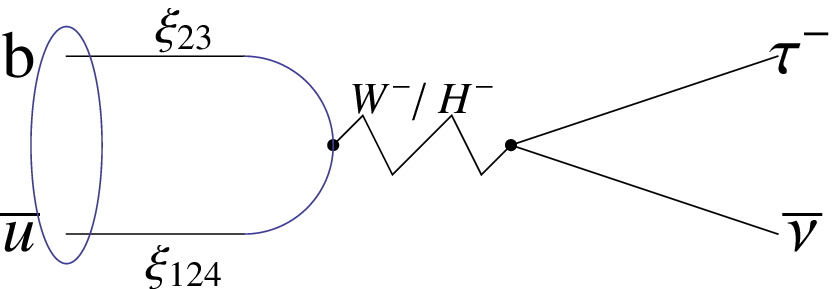}
\caption{Transverse polarization tree diagram 2b.}
\label{Bdecay2b}
\end{center}
\end{minipage}
\end{figure}

\begin{figure}
\begin{minipage}[b]{0.47\linewidth}
\begin{center}
\includegraphics[width=5cm,angle=0,clip]{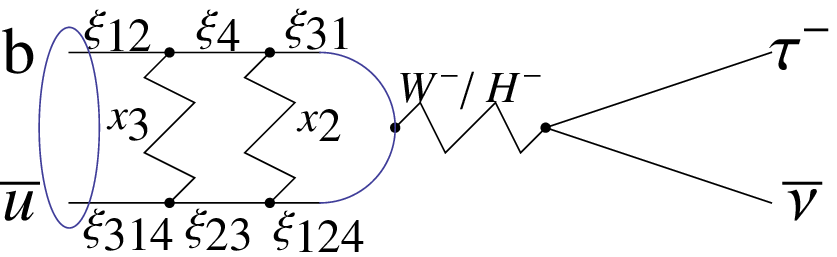}
\caption{The decay of $\bar B$ to $\tau^-\nu$. The penguin diagram 1a.}
\label{penguin1a}
\end{center}
\end{minipage}
\hfill
\begin{minipage}[b]{0.47\linewidth}
\begin{center}
\includegraphics[width=5cm,angle=0,clip]{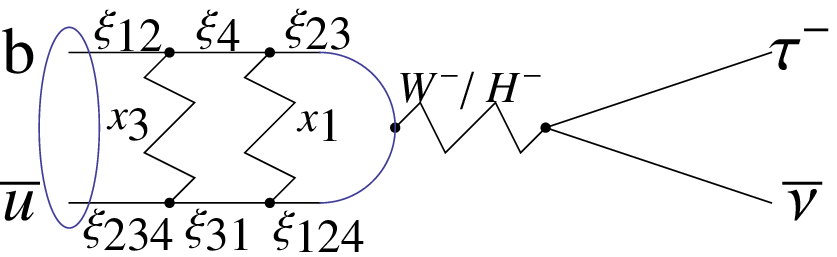}
\caption{The penguin diagram 2a.}
\label{penguin2a}
\end{center}
\end{minipage}
\end{figure}

\begin{figure}
\begin{minipage}[b]{0.47\linewidth}
\begin{center}
\includegraphics[width=5cm,angle=0,clip]{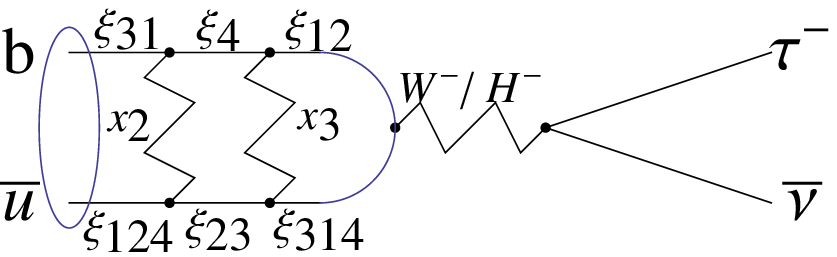}
\caption{The penguin diagram 1b.}
\label{penguin1b}
\end{center}
\end{minipage}
\hfill
\begin{minipage}[b]{0.47\linewidth}
\begin{center}
\includegraphics[width=5cm,angle=0,clip]{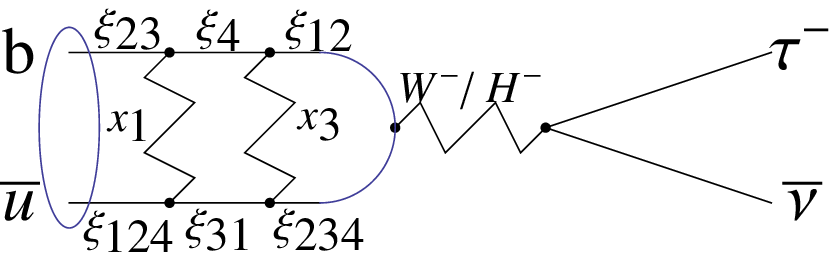}
\caption{The penguin diagram 2b.}
\label{penguin2b}
\end{center}
\end{minipage}
\end{figure}

The model of vertex correction via self-dual gauge fields can be applied to a modification of the SM in weak decay process $B^-\to \tau^-\bar\nu$ \cite{Belle10}, and other meson decays. Some clear deviation from the SM is reported in Penguin diagram dominated processes\cite{LS10}. 
In Figs.\ref{Bdecay1a},\ref{Bdecay2a},\ref{Bdecay1b},\ref{Bdecay2b},  I show the $B^-$ decay into $\tau^-\bar\nu$ in a direct process and in Figs.\ref{penguin1a},\ref{penguin2a},\ref{penguin1b},\ref{penguin2b} in Penguin-like process. Penguin of SM has one lace, but this Penguin has two laces.  In $A_0=0$ gauge and the 3D QCD, the self-dual $W^\pm$ exchange could be evaluatedD

In MSSM the yukawa coupling of $b\bar u$  and $\tau\bar\nu$ to Higgs is given by\cite{Labelle10}
\[
y_u^{ij}(\Psi_{Q_L})_i \tau_2 H^{\dagger T}(\Psi_R)_j = y_u^{bu}(\Psi_{Q_L})_b \tau_2 H^{\dagger T}(\Psi_R)_u
\]
and 
\[
y_e^{ij} (\bar\Psi_L H)_i( \Psi_{e_R})_j = y_e^{ij} (\bar\Psi_L)_i H (\Psi_{e_R})_j,
\]
respectively.
The self-dual gluon exchange contains  infrared divergence and non-perturbative simulation is necessary.

 Georgi\cite{Ge07} assumed that a world with "unparticle" modifies the structure of the infrared fixed point. 
There are investigation of supersymmetry inspired QCD beta function with unparticle incorporated\cite{RS10}. 
If the electron or muon in the detector selects one triality in their electro-weak interactions, most results of SM are not affected. The $q-\bar q$ pair in different triality sector could behave as the unparticle.

In order to explain the mass of W and Z bosons and other fermions, dynamical electro-weak symmetry breaking was proposed by several authors\cite{EL79, ChSi10} with an introduction of technicolor.  
Although the calculation is not done, I guess the flavor changing self-dual gauge field exchange, i.e. $x_1,x_2$ or $x_3$ in Fig.3 and 4 are identified as self-dual gluon or $W^\pm$, affects the mass of gauge bosons.

It is necessary to extend the lattice simulation to larger lattice and study the continuum limit.

\leftline{\bf Acknowledgement}
I thank Stan Brodsky for helpful suggestions.
The DWF lattice simulations related to this work were performed at SR8000 at KEK High Energy Accelerator Research Organization, SX5 at RCNP Osaka university, SR16000 at Yukawa institute for fundamental physics and T2K at Tsukuba university.
  
\newpage
\vskip 0.5 true cm
{\bf Appendix: Matrix elements of the triality transformation of the $g_2$ algebra.}
\vskip 0.3 true cm
In the table $\bar 1$ indicates -1.
In $G_{12}$, I replaced $\xi_4\to -\xi_{1234}$ to $\xi_4\to -\xi_{123}$ written in \cite{Cartan66} (probably typos). 
$G_{13}G_{12}$ does not agree with $G_{(123)}$ of \cite{Cartan66}. We think
$\xi_4\to -x^{4'}$ there should be replaced by $\xi_4\to -x^4$ and we define
${^t(G}_{12})G_{13}\equiv {G}_{123}$ and ${^t(G}_{13})G_{12}\equiv{G}_{132}$, which correspond to $G_{(123)}$ and $G_{(132)}$ respectively of \cite{Cartan66}.

$G_{123}^{-1}-G_{132}=-2\delta_{\xi_0 x_4'}$ and $G_{132}^{-1}-G_{123}=2\delta_{x_4'\xi_0}$, i.e. inversion is same as interchange except the sign of $\xi_0\to x_4'$ transformation.

 Under these triality automorphism, the set $\{F,\Phi,\Psi\}$ is invariant.

\begin{landscape}
${\small
\begin{array}{ccccccccccccccccccccccccc}
G_{23}&\xi_0& \xi_1& \xi_2& \xi_3& \xi_4&\xi_{12}& \xi_{31}& \xi_{23}& \xi_{14}& \xi_{24}& \xi_{34}&\xi_{123}& \xi_{124}& \xi_{314}& \xi_{234}& \xi_{1234}&x^1& x^2& x^3& x^4&x^{1'}& x^{2'}& x^{3'}& x^{4'}\\
 \xi_0&0 & 0 & 0 & 0 & 1 & 0 & 0 & 0 & 0 & 0 & 0 & 0 & 0 & 0 & 0 & 0 & 0 & 0 & 0 & 0 & 0 & 0 & 0 & 0 \\
 \xi_1&0 & 0 & 0 & 0 & 0 & 0 & 0 & 0 & \bar 1 & 0 & 0 & 0 & 0 & 0 & 0 & 0 & 0 & 0 & 0 & 0 & 0 & 0 & 0 & 0 \\
 \xi_2&0 & 0 & 0 & 0 & 0 & 0 & 0 & 0 & 0 & \bar 1 & 0 & 0 & 0 & 0 & 0 & 0 & 0 & 0 & 0 & 0 & 0 & 0 & 0 & 0 \\
\xi_3& 0 & 0 & 0 & 0 & 0 & 0 & 0 & 0 & 0 & 0 & \bar 1 & 0 & 0 & 0 & 0 & 0 & 0 & 0 & 0 & 0 & 0 & 0 & 0 & 0 \\
\xi_4& \bar 1 & 0 & 0 & 0 & 0 & 0 & 0 & 0 & 0 & 0 & 0 & 0 & 0 & 0 & 0 & 0 & 0 & 0 & 0 & 0 & 0 & 0 & 0 & 0 \\
\xi_{12}& 0 & 0 & 0 & 0 & 0 & 0 & 0 & 0 & 0 & 0 & 0 & 0 & 1 & 0 & 0 & 0 & 0 & 0 & 0 & 0 & 0 & 0 & 0 & 0 \\
\xi_{31}& 0 & 0 & 0 & 0 & 0 & 0 & 0 & 0 & 0 & 0 & 0 & 0 & 0 & 1 & 0 & 0 & 0 & 0 & 0 & 0 & 0 & 0 & 0 & 0 \\
\xi_{23}& 0 & 0 & 0 & 0 & 0 & 0 & 0 & 0 & 0 & 0 & 0 & 0 & 0 & 0 & 1 & 0 & 0 & 0 & 0 & 0 & 0 & 0 & 0 & 0 \\
\xi_{14}& 0 & 1 & 0 & 0 & 0 & 0 & 0 & 0 & 0 & 0 & 0 & 0 & 0 & 0 & 0 & 0 & 0 & 0 & 0 & 0 & 0 & 0 & 0 & 0 \\
\xi_{24}& 0 & 0 & 1 & 0 & 0 & 0 & 0 & 0 & 0 & 0 & 0 & 0 & 0 & 0 & 0 & 0 & 0 & 0 & 0 & 0 & 0 & 0 & 0 & 0 \\
\xi_{34}& 0 & 0 & 0 & 1 & 0 & 0 & 0 & 0 & 0 & 0 & 0 & 0 & 0 & 0 & 0 & 0 & 0 & 0 & 0 & 0 & 0 & 0 & 0 & 0 \\
\xi_{123}& 0 & 0 & 0 & 0 & 0 & 0 & 0 & 0 & 0 & 0 & 0 & 0 & 0 & 0 & 0 & \bar 1 & 0 & 0 & 0 & 0 & 0 & 0 & 0 & 0 \\
\xi_{124}& 0 & 0 & 0 & 0 & 0 & \bar 1 & 0 & 0 & 0 & 0 & 0 & 0 & 0 & 0 & 0 & 0 & 0 & 0 & 0 & 0 & 0 & 0 & 0 & 0 \\
\xi_{314}& 0 & 0 & 0 & 0 & 0 & 0 & \bar 1 & 0 & 0 & 0 & 0 & 0 & 0 & 0 & 0 & 0 & 0 & 0 & 0 & 0 & 0 & 0 & 0 & 0 \\
\xi_{234}& 0 & 0 & 0 & 0 & 0 & 0 & 0 & \bar 1 & 0 & 0 & 0 & 0 & 0 & 0 & 0 & 0 & 0 & 0 & 0 & 0 & 0 & 0 & 0 & 0 \\
\xi_{1234}& 0 & 0 & 0 & 0 & 0 & 0 & 0 & 0 & 0 & 0 & 0 & 1 & 0 & 0 & 0 & 0 & 0 & 0 & 0 & 0 & 0 & 0 & 0 & 0 \\
x^1 & 0 & 0 & 0 & 0 & 0 & 0 & 0 & 0 & 0 & 0 & 0 & 0 & 0 & 0 & 0 & 0 & 1 & 0 & 0 & 0 & 0 & 0 & 0 & 0 \\
x^2& 0 & 0 & 0 & 0 & 0 & 0 & 0 & 0 & 0 & 0 & 0 & 0 & 0 & 0 & 0 & 0 & 0 & 1 & 0 & 0 & 0 & 0 & 0 & 0 \\
x^3& 0 & 0 & 0 & 0 & 0 & 0 & 0 & 0 & 0 & 0 & 0 & 0 & 0 & 0 & 0 & 0 & 0 & 0 & 1 & 0 & 0 & 0 & 0 & 0 \\
x^4& 0 & 0 & 0 & 0 & 0 & 0 & 0 & 0 & 0 & 0 & 0 & 0 & 0 & 0 & 0 & 0 & 0 & 0 & 0 & 0 & 0 & 0 & 0 & \bar 1 \\
x^{1'}& 0 & 0 & 0 & 0 & 0 & 0 & 0 & 0 & 0 & 0 & 0 & 0 & 0 & 0 & 0 & 0 & 0 & 0 & 0 & 0 & 1 & 0 & 0 & 0 \\
x^{2'}& 0 & 0 & 0 & 0 & 0 & 0 & 0 & 0 & 0 & 0 & 0 & 0 & 0 & 0 & 0 & 0 & 0 & 0 & 0 & 0 & 0 & 1 & 0 & 0 \\
x^{3'}& 0 & 0 & 0 & 0 & 0 & 0 & 0 & 0 & 0 & 0 & 0 & 0 & 0 & 0 & 0 & 0 & 0 & 0 & 0 & 0 & 0 & 0 & 1 & 0 \\
x^{4'}& 0 & 0 & 0 & 0 & 0 & 0 & 0 & 0 & 0 & 0 & 0 & 0 & 0 & 0 & 0 & 0 & 0 & 0 & 0 & \bar 1 & 0 & 0 & 0 & 0
\end{array}
}$

\newpage
${\small
\begin{array}{ccccccccccccccccccccccccc}
G_{12} &\xi_0& \xi_1& \xi_2& \xi_3& \xi_4&\xi_{12}& \xi_{31}& \xi_{23}& \xi_{14}& \xi_{24}& \xi_{34}&\xi_{123}& \xi_{124}& \xi_{314}& \xi_{234}& \xi_{1234}&x^1& x^2& x^3& x^4&x^{1'}& x^{2'}& x^{3'}& x^{4'}\\
\xi_0& 0 & 0 & 0 & 0 & 0 & 0 & 0 & 0 & 0 & 0 & 0 & 0 & 0 & 0 & 0 & 0 & 0 & 0 & 0 & 1 & 0 & 0 & 0 & 0 \\
\xi_1& 0 & 1 & 0 & 0 & 0 & 0 & 0 & 0 & 0 & 0 & 0 & 0 & 0 & 0 & 0 & 0 & 0 & 0 & 0 & 0 & 0 & 0 & 0 & 0 \\
\xi_2& 0 & 0 & 1 & 0 & 0 & 0 & 0 & 0 & 0 & 0 & 0 & 0 & 0 & 0 & 0 & 0 & 0 & 0 & 0 & 0 & 0 & 0 & 0 & 0 \\
\xi_3& 0 & 0 & 0 & 1 & 0 & 0 & 0 & 0 & 0 & 0 & 0 & 0 & 0 & 0 & 0 & 0 & 0 & 0 & 0 & 0 & 0 & 0 & 0 & 0 \\
\xi_4& 0 & 0 & 0 & 0 & 0 & 0 & 0 & 0 & 0 & 0 & 0 & \bar 1 & 0 & 0 & 0 & 0 & 0 & 0 & 0 & 0 & 0 & 0 & 0 & 0 \\
\xi_{12}& 0 & 0 & 0 & 0 & 0 & 0 & 0 & 0 & 0 & 0 & 0 & 0 & 0 & 0 & 0 & 0 & 0 & 0 & 1 & 0 & 0 & 0 & 0 & 0 \\
\xi_{31}& 0 & 0 & 0 & 0 & 0 & 0 & 0 & 0 & 0 & 0 & 0 & 0 & 0 & 0 & 0 & 0 & 0 & 1 & 0 & 0 & 0 & 0 & 0 & 0 \\
\xi_{23}& 0 & 0 & 0 & 0 & 0 & 0 & 0 & 0 & 0 & 0 & 0 & 0 & 0 & 0 & 0 & 0 & 1 & 0 & 0 & 0 & 0 & 0 & 0 & 0 \\
\xi_{14}& 0 & 0 & 0 & 0 & 0 & 0 & 0 & 0 & 0 & 0 & 0 & 0 & 0 & 0 & 0 & 0 & 0 & 0 & 0 & 0 & \bar 1 & 0 & 0 & 0 \\
\xi_{24}& 0 & 0 & 0 & 0 & 0 & 0 & 0 & 0 & 0 & 0 & 0 & 0 & 0 & 0 & 0 & 0 & 0 & 0 & 0 & 0 & 0 & \bar 1 & 0 & 0 \\
\xi_{34}& 0 & 0 & 0 & 0 & 0 & 0 & 0 & 0 & 0 & 0 & 0 & 0 & 0 & 0 & 0 & 0 & 0 & 0 & 0 & 0 & 0 & 0 & \bar 1 & 0 \\
\xi_{123}& 0 & 0 & 0 & 0 & \bar 1 & 0 & 0 & 0 & 0 & 0 & 0 & 0 & 0 & 0 & 0 & 0 & 0 & 0 & 0 & 0 & 0 & 0 & 0 & 0 \\
\xi_{124}& 0 & 0 & 0 & 0 & 0 & 0 & 0 & 0 & 0 & 0 & 0 & 0 & 1 & 0 & 0 & 0 & 0 & 0 & 0 & 0 & 0 & 0 & 0 & 0 \\
\xi_{314}& 0 & 0 & 0 & 0 & 0 & 0 & 0 & 0 & 0 & 0 & 0 & 0 & 0 & 1 & 0 & 0 & 0 & 0 & 0 & 0 & 0 & 0 & 0 & 0 \\
\xi_{234}& 0 & 0 & 0 & 0 & 0 & 0 & 0 & 0 & 0 & 0 & 0 & 0 & 0 & 0 & 1 & 0 & 0 & 0 & 0 & 0 & 0 & 0 & 0 & 0 \\
\xi_{1234}& 0 & 0 & 0 & 0 & 0 & 0 & 0 & 0 & 0 & 0 & 0 & 0 & 0 & 0 & 0 & 0 & 0 & 0 & 0 & 0 & 0 & 0 & 0 & 1 \\
x^1& 0 & 0 & 0 & 0 & 0 & 0 & 0 & \bar 1 & 0 & 0 & 0 & 0 & 0 & 0 & 0 & 0 & 0 & 0 & 0 & 0 & 0 & 0 & 0 & 0 \\
x^2& 0 & 0 & 0 & 0 & 0 & 0 & \bar 1 & 0 & 0 & 0 & 0 & 0 & 0 & 0 & 0 & 0 & 0 & 0 & 0 & 0 & 0 & 0 & 0 & 0 \\
x^3& 0 & 0 & 0 & 0 & 0 & \bar 1 & 0 & 0 & 0 & 0 & 0 & 0 & 0 & 0 & 0 & 0 & 0 & 0 & 0 & 0 & 0 & 0 & 0 & 0 \\
x^4& 1 & 0 & 0 & 0 & 0 & 0 & 0 & 0 & 0 & 0 & 0 & 0 & 0 & 0 & 0 & 0 & 0 & 0 & 0 & 0 & 0 & 0 & 0 & 0 \\
x^{1'}& 0 & 0 & 0 & 0 & 0 & 0 & 0 & 0 & \bar 1 & 0 & 0 & 0 & 0 & 0 & 0 & 0 & 0 & 0 & 0 & 0 & 0 & 0 & 0 & 0 \\
x^{2'}& 0 & 0 & 0 & 0 & 0 & 0 & 0 & 0 & 0 & \bar 1 & 0 & 0 & 0 & 0 & 0 & 0 & 0 & 0 & 0 & 0 & 0 & 0 & 0 & 0 \\
x^{3'}& 0 & 0 & 0 & 0 & 0 & 0 & 0 & 0 & 0 & 0 & \bar 1 & 0 & 0 & 0 & 0 & 0 & 0 & 0 & 0 & 0 & 0 & 0 & 0 & 0 \\
x^{4'}& 0 & 0 & 0 & 0 & 0 & 0 & 0 & 0 & 0 & 0 & 0 & 0 & 0 & 0 & 0 & \bar 1 & 0 & 0 & 0 & 0 & 0 & 0 & 0 & 0
\end{array}
}$
\newpage
${\small
\begin{array}{cccccccccccccccccccccccccc}
G_{13}&\xi_0& \xi_1& \xi_2& \xi_3& \xi_4&\xi_{12}& \xi_{31}& \xi_{23}& \xi_{14}& \xi_{24}& \xi_{34}&\xi_{123}& \xi_{124}& \xi_{314}& \xi_{234}& \xi_{1234}&x^1& x^2& x^3& x^4&x^{1'}& x^{2'}& x^{3'}& x^{4'}\\
\xi_0 & 0 & 0 & 0 & 0 & 0 & 0 & 0 & 0 & 0 & 0 & 0 & 0 & 0 & 0 & 0 & \bar 1 & 0 & 0 & 0 & 0 & 0 & 0 & 0 & 0 \\
\xi_1& 0 & 0 & 0 & 0 & 0 & 0 & 0 & 0 & 0 & 0 & 0 & 0 & 0 & 0 & 0 & 0 & 0 & 0 & 0 & 0 & 1 & 0 & 0 & 0 \\
\xi_2& 0 & 0 & 0 & 0 & 0 & 0 & 0 & 0 & 0 & 0 & 0 & 0 & 0 & 0 & 0 & 0 & 0 & 0 & 0 & 0 & 0 & 1 & 0 & 0 \\
\xi_3& 0 & 0 & 0 & 0 & 0 & 0 & 0 & 0 & 0 & 0 & 0 & 0 & 0 & 0 & 0 & 0 & 0 & 0 & 0 & 0 & 0 & 0 & 1 & 0 \\
\xi_4& 0 & 0 & 0 & 0 & 0 & 0 & 0 & 0 & 0 & 0 & 0 & 0 & 0 & 0 & 0 & 0 & 0 & 0 & 0 & 0 & 0 & 0 & 0 & 1 \\
\xi_{12}& 0 & 0 & 0 & 0 & 0 & 1 & 0 & 0 & 0 & 0 & 0 & 0 & 0 & 0 & 0 & 0 & 0 & 0 & 0 & 0 & 0 & 0 & 0 & 0 \\
\xi_{31}& 0 & 0 & 0 & 0 & 0 & 0 & 1 & 0 & 0 & 0 & 0 & 0 & 0 & 0 & 0 & 0 & 0 & 0 & 0 & 0 & 0 & 0 & 0 & 0 \\
\xi_{23}& 0 & 0 & 0 & 0 & 0 & 0 & 0 & 1 & 0 & 0 & 0 & 0 & 0 & 0 & 0 & 0 & 0 & 0 & 0 & 0 & 0 & 0 & 0 & 0 \\
\xi_{14}& 0 & 0 & 0 & 0 & 0 & 0 & 0 & 0 & 1 & 0 & 0 & 0 & 0 & 0 & 0 & 0 & 0 & 0 & 0 & 0 & 0 & 0 & 0 & 0 \\
\xi_{24}& 0 & 0 & 0 & 0 & 0 & 0 & 0 & 0 & 0 & 1 & 0 & 0 & 0 & 0 & 0 & 0 & 0 & 0 & 0 & 0 & 0 & 0 & 0 & 0 \\
\xi_{34}& 0 & 0 & 0 & 0 & 0 & 0 & 0 & 0 & 0 & 0 & 1 & 0 & 0 & 0 & 0 & 0 & 0 & 0 & 0 & 0 & 0 & 0 & 0 & 0 \\
\xi_{123}& 0 & 0 & 0 & 0 & 0 & 0 & 0 & 0 & 0 & 0 & 0 & 0 & 0 & 0 & 0 & 0 & 0 & 0 & 0 & 1 & 0 & 0 & 0 & 0 \\
\xi_{124}& 0 & 0 & 0 & 0 & 0 & 0 & 0 & 0 & 0 & 0 & 0 & 0 & 0 & 0 & 0 & 0 & 0 & 0 & \bar 1 & 0 & 0 & 0 & 0 & 0 \\
\xi_{314}& 0 & 0 & 0 & 0 & 0 & 0 & 0 & 0 & 0 & 0 & 0 & 0 & 0 & 0 & 0 & 0 & 0 & \bar 1 & 0 & 0 & 0 & 0 & 0 & 0 \\
\xi_{234}& 0 & 0 & 0 & 0 & 0 & 0 & 0 & 0 & 0 & 0 & 0 & 0 & 0 & 0 & 0 & 0 & \bar 1 & 0 & 0 & 0 & 0 & 0 & 0 & 0 \\
\xi_{1234}& \bar 1 & 0 & 0 & 0 & 0 & 0 & 0 & 0 & 0 & 0 & 0 & 0 & 0 & 0 & 0 & 0 & 0 & 0 & 0 & 0 & 0 & 0 & 0 & 0 \\
x^1& 0 & 0 & 0 & 0 & 0 & 0 & 0 & 0 & 0 & 0 & 0 & 0 & 0 & 0 & 1 & 0 & 0 & 0 & 0 & 0 & 0 & 0 & 0 & 0 \\
x^2& 0 & 0 & 0 & 0 & 0 & 0 & 0 & 0 & 0 & 0 & 0 & 0 & 0 & 1 & 0 & 0 & 0 & 0 & 0 & 0 & 0 & 0 & 0 & 0 \\
x^3& 0 & 0 & 0 & 0 & 0 & 0 & 0 & 0 & 0 & 0 & 0 & 0 & 1 & 0 & 0 & 0 & 0 & 0 & 0 & 0 & 0 & 0 & 0 & 0 \\
x^4& 0 & 0 & 0 & 0 & 0 & 0 & 0 & 0 & 0 & 0 & 0 & \bar 1 & 0 & 0 & 0 & 0 & 0 & 0 & 0 & 0 & 0 & 0 & 0 & 0 \\
x^{1'}& 0 & \bar 1 & 0 & 0 & 0 & 0 & 0 & 0 & 0 & 0 & 0 & 0 & 0 & 0 & 0 & 0 & 0 & 0 & 0 & 0 & 0 & 0 & 0 & 0 \\
x^{2'}& 0 & 0 & \bar 1 & 0 & 0 & 0 & 0 & 0 & 0 & 0 & 0 & 0 & 0 & 0 & 0 & 0 & 0 & 0 & 0 & 0 & 0 & 0 & 0 & 0 \\
x^{3'}& 0 & 0 & 0 & \bar 1 & 0 & 0 & 0 & 0 & 0 & 0 & 0 & 0 & 0 & 0 & 0 & 0 & 0 & 0 & 0 & 0 & 0 & 0 & 0 & 0 \\
x^{4'}& 0 & 0 & 0 & 0 & \bar 1 & 0 & 0 & 0 & 0 & 0 & 0 & 0 & 0 & 0 & 0 & 0 & 0 & 0 & 0 & 0 & 0 & 0 & 0 & 0
\end{array}
}$
\newpage
${\small
\begin{array}{cccccccccccccccccccccccccc}
G_{123}&\xi_0& \xi_1& \xi_2& \xi_3& \xi_4&\xi_{12}& \xi_{31}& \xi_{23}& \xi_{14}& \xi_{24}& \xi_{34}&\xi_{123}& \xi_{124}& \xi_{314}& \xi_{234}& \xi_{1234}&x^1& x^2& x^3& x^4&x^{1'}& x^{2'}& x^{3'}& x^{4'}\\
\xi_0&  0 &0 &0 &0 &0 &0 &0 &0 &0 &0 &0 &\bar 1 &0 &0 &0 &0 &0 &0 &0 &0 &0 &0 &0 &0\\
\xi_1&  0 &0 &0 &0 &0 &0 &0 &0 &0 &0 &0 &0 &0 &0 &0 &0 &0 &0 &0 &0 & 1 &0 &0 &0\\ 
\xi_2& 0 &0 &0 &0 &0 &0 &0 &0 &0 &0 &0 &0 &0 &0 &0 &0 &0 &0 &0 &0 &0 & 1 &0 &0\\ 
\xi_3& 0 &0 &0 &0 &0 &0 &0 &0 &0 &0 &0 &0 &0 &0 &0 &0 &0 &0 &0 &0 &0 &0 & 1 &0\\
\xi_4&  0 &0 &0 &0 &0 &0 &0 &0 &0 &0 &0 &0 &0 &0 &0 &0 &0 &0 &0 &\bar 1 &0 &0 &0 &0\\ 
\xi_{12}&  0 &0 &0 &0 &0 &0 &0 &0 &0 &0 &0 &0 &\bar 1 &0 &0 &0 &0 &0 &0 &0 &0 &0 &0 &0\\ 
\xi_{31}&  0 &0 &0 &0 &0 &0 &0 &0 &0 &0 &0 &0 &0 &\bar 1 &0 &0 &0 &0 &0 &0 &0 &0 &0 &0\\ 
\xi_{23}&  0 &0 &0 &0 &0 &0 &0 &0 &0 &0 &0 &0 &0 &0 &\bar 1 &0 &0 &0 &0 &0 &0 &0 &0 &0\\ 
\xi_{14}& 0 & 1 &0 &0 &0 &0 &0 &0 &0 &0 &0 &0 &0 &0 &0 &0 &0 &0 &0 &0 &0 &0 &0 &0\\ 
\xi_{24}& 0 &0 & 1 &0 &0 &0 &0 &0 &0 &0 &0 &0 &0 &0 &0 &0 &0 &0 &0 &0 &0 &0 &0 &0\\ 
\xi_{34}& 0 &0 &0 & 1 &0 &0 &0 &0 &0 &0 &0 &0 &0 &0 &0 &0 &0 &0 &0 &0 &0 &0 &0 &0\\ 
\xi_{123}&  0 &0 &0 &0 &0 &0 &0 &0 &0 &0 &0 &0 &0 &0 &0 &0 &0 &0 &0 &0 &0 &0 &0 &\bar 1\\ 
\xi_{124}&  0 &0 &0 &0 &0 &0 &0 &0 &0 &0 &0 &0 &0 &0 &0 &0 &0 &0 &\bar 1 &0 &0 &0 &0 &0\\ 
\xi_{314}&  0 &0 &0 &0 &0 &0 &0 &0 &0 &0 &0 &0 &0 &0 &0 &0 &0 &\bar 1 &0 &0 &0 &0 &0 &0\\ 
\xi_{234}&  0 &0 &0 &0 &0 &0 &0 &0 &0 &0 &0 &0 &0 &0 &0 &0 &\bar 1 &0 &0 &0 &0 &0 &0 &0\\ 
\xi_{1234}& 0 &0 &0 &0 &1 &0 &0 &0 &0 &0 &0 &0 &0 &0 &0 &0 &0 &0 &0 &0 &0 &0 &0 &0\\ 
x^1&  0 &0 &0 &0 &0 &0 &0 &1 &0 &0 &0 &0 &0 &0 &0 &0 &0 &0 &0 &0 &0 &0 &0 &0\\ 
x^2&  0 &0 &0 &0 &0 &0 &1 &0 &0 &0 &0 &0 &0 &0 &0 &0 &0 &0 &0 &0 &0 &0 &0 &0\\ 
x^3&  0 &0 &0 &0 &0 &1 &0 &0 &0 &0 &0 &0 &0 &0 &0 &0 &0 &0 &0 &0 &0 &0 &0 &0\\
x^4&  0 &0 &0 &0 &0 &0 &0 &0 &0 &0 &0 &0 &0 &0 &0 &\bar 1 &0 &0 &0 &0 &0 &0 &0 &0\\ 
x^{1'}& 0 &0 &0 &0 &0 &0 &0 &0 &\bar 1 &0 &0 &0 &0 &0 &0 &0 &0 &0 &0 &0 &0 &0 &0 &0\\ 
x^{2'}& 0 &0 &0 &0 &0 &0 &0 &0 &0 &\bar 1 &0 &0 &0 &0 &0 &0 &0 &0 &0 &0 &0 &0 &0 &0\\ 
x^{3'}& 0 &0 &0 &0 &0 &0 &0 &0 &0 &0 &\bar 1 &0 &0 &0 &0 &0 &0 &0 &0 &0 &0 &0 &0 &0\\ 
x^{4'}& \bar 1 &0 &0 &0 &0 &0 &0 &0 &0 &0 &0 &0 &0 &0 &0 &0 &0 &0 &0 &0 &0 &0 &0 &0\\
\end{array}
}$
\newpage
${\small
\begin{array}{cccccccccccccccccccccccccc}
G_{132}&\xi_0& \xi_1& \xi_2& \xi_3& \xi_4&\xi_{12}& \xi_{31}& \xi_{23}& \xi_{14}& \xi_{24}& \xi_{34}&\xi_{123}& \xi_{124}& \xi_{314}& \xi_{234}& \xi_{1234}&x^1& x^2& x^3& x^4&x^{1'}& x^{2'}& x^{3'}& x^{4'}\\
\xi_0& 0 &0 &0 &0 &0 &0 &0 &0 &0 &0 &0 &0 &0 &0 &0 &0 &0 &0 &0 &0 &0 &0 &0 &\bar  1\\ 
\xi_1&  0 &0 &0 &0 &0 &0 &0 &0 &1 &0 &0 &0 &0 &0 &0 &0 &0 &0 &0 &0 &0 &0 &0 &0\\ 
\xi_2&  0 &0 &0 &0 &0 &0 &0 &0 &0 &\bar 1 &0 &0 &0 &0 &0 &0 &0 &0 &0 &0 &0 &0 &0 &0\\ 
\xi_3&  0 &0 &0 &0 &0 &0 &0 &0 &0 &0 &\bar 1 &0 &0 &0 &0 &0 &0 &0 &0 &0 &0 &0 &0 &0\\
\xi_4&  0 &0 &0 &0 &0 &0 &0 &0 &0 &0 &0 &0 &0 &0 &0 &1 &0 &0 &0 &0 &0 &0 &0 &0\\
\xi_{12}&  0 &0 &0 &0 &0 &0 &0 &0 &0 &0 &0 &0 &0 &0 &0 &0 &0 &0 & 1 &0 &0 &0 &0 &0\\ 
\xi_{31}& 0 &0 &0 &0 &0 &0 &0 &0 &0 &0 &0 &0 &0 &0 &0 &0 &0 & 1 &0 &0 &0 &0 &0 &0\\
\xi_{23}&  0 &0 &0 &0 &0 &0 &0 &0 &0 &0 &0 &0 &0 &0 &0 &0 & 1 &0 &0 &0 &0 &0 &0 &0\\ 
\xi_{14}& 0 &0 &0 &0 &0 &0 &0 &0 &0 &0 &0 &0 &0 &0 &0 &0 &0 &0 &0 &0 &\bar 1 &0 &0 &0\\
\xi_{24}&  0 &0 &0 &0 &0 &0 &0 &0 &0 &0 &0 &0 &0 &0 &0 &0 &0 &0 &0 &0 &0 &\bar 1 &0 &0\\
\xi_{34}&  0 &0 &0 &0 &0 &0 &0 &0 &0 &0 &0 &0 &0 &0 &0 &0 &0 &0 &0 &0 &0 &0 &\bar 1 &0\\ 
\xi_{123}&\bar 1 &0 &0 &0 &0 &0 &0 &0 &0 &0 &0 &0 &0 &0 &0 &0 &0 &0 &0 &0 &0 &0 &0 &0\\ 
\xi_{124}&  0 &0 &0 &0 &0 &\bar 1 &0 &0 &0 &0 &0 &0 &0 &0 &0 &0 &0 &0 &0 &0 &0 &0 &0 &0\\ 
\xi_{314}&  0 &0 &0 &0 &0 &0 &\bar 1 &0 &0 &0 &0 &0 &0 &0 &0 &0 &0 &0 &0 &0 &0 &0 &0 &0\\ 
\xi_{234}&  0 &0 &0 &0 &0 &0 &0 &\bar 1 &0 &0 &0 &0 &0 &0 &0 &0 &0 &0 &0 &0 &0 &0 &0 &0\\ 
\xi_{1234}&  0 &0 &0 &0 &0 &0 &0 &0 &0 &0 &0 &0 &0 &0 &0 &0 &0 &0 &0 &\bar 1 &0 &0 &0 &0\\
x^1&  0 &0 &0 &0 &0 &0 &0 &0 &0 &0 &0 &0 &0 &0 &\bar 1 &0 &0 &0 &0 &0 &0 &0 &0 &0\\ 
x^2& 0 &0 &0 &0 &0 &0 &0 &0 &0 &0 &0 &0 &0 &\bar 1 &0 &0 &0 &0 &0 &0 &0 &0 &0 &0\\ 
x^3& 0 &0 &0 &0 &0 &0 &0 &0 &0 &0 &0 &0 &\bar 1 &0 &0 &0 &0 &0 &0 &0 &0 &0 &0 &0\\ 
x^4& 0 &0 &0 &0 &\bar 1 &0 &0 &0 &0 &0 &0 &0 &0 &0 &0 &0 &0 &0 &0 &0 &0 &0 &0 &0\\ 
x^{1'}&  0 &1 &0 &0 &0 &0 &0 &0 &0 &0 &0 &0 &0 &0 &0 &0 &0 &0 &0 &0 &0 &0 &0 &0\\ 
x^{2'}&  0 &0 &1 &0 &0 &0 &0 &0 &0 &0 &0 &0 &0 &0 &0 &0 &0 &0 &0 &0 &0 &0 &0 &0\\ 
x^{3'}&  0 &0 &0 &1 &0 &0 &0 &0 &0 &0 &0 &0 &0 &0 &0 &0 &0 &0 &0 &0 &0 &0 &0 &0\\ 
x^{4'}& 0 &0 &0 &0 &0 &0 &0 &0 &0 &0 &0 &\bar 1 &0 &0 &0 &0 &0 &0 &0 &0 &0 &0 &0 &0\\
\end{array}
}$
\end{landscape}
\end{document}